\documentclass{aa}

\usepackage[utf8]{inputenc}
\usepackage[T1]{fontenc}
\usepackage{xspace}
\usepackage{float}
\usepackage{color}
\usepackage[dvipsnames]{xcolor}
\usepackage{graphicx}
\usepackage{pgf}
\graphicspath{{gfx/}}
\usepackage{subcaption}
\usepackage{booktabs}
\usepackage{lmodern}
\usepackage{import}

\everymath=\expandafter{\the\everymath\displaystyle}
\usepackage{amsmath}
\usepackage[normalem]{ulem} 

\usepackage{amssymb}
\usepackage{listings}

\definecolor{codegreen}{rgb}{0,0.6,0}
\definecolor{codegray}{rgb}{0.5,0.5,0.5}
\definecolor{codepurple}{rgb}{0.58,0,0.82}
\definecolor{backcolour}{rgb}{0.95,0.95,0.92}

\lstdefinestyle{mystyle}{
    backgroundcolor=\color{backcolour},   
    commentstyle=\color{codegreen},
    keywordstyle=\color{magenta},
    numberstyle=\tiny\color{codegray},
    stringstyle=\color{codepurple},
    basicstyle=\ttfamily\footnotesize,
    breakatwhitespace=false,
    breaklines=true,
    captionpos=b,
    keepspaces=true,
    numbers=left,
    numbersep=5pt,
    showspaces=false,
    showstringspaces=false,
    showtabs=false,
    tabsize=2
}
\usepackage{aux/aas_macros}
\usepackage{natbib}
\bibpunct{(}{)}{;}{a}{}{,} 

\usepackage{url}
\usepackage[colorlinks=True,
  linkcolor=gray,         
  citecolor=RoyalBlue,    
  filecolor=gray,         
  urlcolor=gray           
]{hyperref}

\usepackage[nameinlink,capitalize]{cleveref}
\crefformat{equation}{Eq.~#2#1#3}
\crefname{section}{Sect.}{Sects.} 

\usepackage{lipsum}
\usepackage[switch]{lineno}

\usepackage[nameinlink]{cleveref}

\usepackage{catoptions}

\makeatletter

\def\Autoref#1{%
  \begingroup
  \edef\reserved@a{\cpttrimspaces{#1}}%
  \ifcsndefTF{r@#1}{%
    \xaftercsname{\expandafter\testreftype\@fourthoffive}
      {r@\reserved@a}.\\{#1}%
  }{%
    \ref{#1}%
  }%
  \endgroup
}
\def\testreftype#1.#2\\#3{%
  \ifcsndefTF{#1autorefname}{%
    \def\reserved@a##1##2\@nil{%
      \uppercase{\def\ref@name{##1}}%
      \csn@edef{#1autorefname}{\ref@name##2}%
      \autoref{#3}%
    }%
    \reserved@a#1\@nil
  }{%
    \autoref{#3}%
  }%
}

\makeatother

\newcommand{\rev}[1]{\textcolor{orange}{\bf #1}}
\newcommand{\del}[1]{\textcolor{red}{\sout{#1}}}

\renewcommand{\rev}[1]{#1}
\renewcommand{\del}[1]{}

\newcommand{\miriade}{\texttt{Miriade}\xspace}
\newcommand{\ssodnet}{\texttt{SsODNet}\xspace}

\newcommand{\rocks}{\texttt{rocks}\xspace}

\newcommand{\topcat}{\texttt{TOPCAT}\xspace}
\newcommand{\astropy}{\texttt{astropy}\xspace}
\newcommand{\sbpy}{\texttt{sbpy}\xspace}
\newcommand{\sorcha}{\texttt{SORCHA}\xspace}
\newcommand{\nifty}{\texttt{nifty-ls}\xspace}

\newcommand{\fink}{\textsc{Fink}\xspace}

\newcommand{\ztf}{\textsc{ZTF}\xspace}

\newcommand{\lsst}{\textsc{LSST}\xspace}

\renewcommand{\degr}{\ensuremath{^{\textrm{o}}}\xspace}

\newcommand{\funcf}{\ensuremath{f(r,\Delta)}\xspace}
\newcommand{\funcg}{\ensuremath{g(\gamma)}\xspace}
\newcommand{\funcs}{\ensuremath{s(\alpha,\delta)}\xspace}
\newcommand{\funcss}{\ensuremath{s(\alpha,\delta,t)}\xspace}
\newcommand{\hg}{\texttt{HG}\xspace}
\newcommand{\hgs}{\texttt{HG$_{12}^{\star}$}\xspace}
\newcommand{\hgg}{\texttt{HG$_1$G$_2$}\xspace}

\newcommand{\shgg}{\texttt{sHG$_1$G$_2$}\xspace}

\newcommand{\socca}{\texttt{SOCCA}\xspace}
\newcommand{\soccaname}{Shape, Orientation and Colors Combined approach for Asteroids\xspace}

\newcommand{\nfilter}{\ensuremath{N_{F}}\xspace}
\newcommand{\nobs}{\ensuremath{N_{\textrm{o}}}\xspace}

\newcommand{\ggparams}{\texttt{G$_1$G$_2$}\xspace}
\newcommand{\Hmag}{\texttt{H}\xspace}
\newcommand{\gone}{\texttt{G$_1$}\xspace}
\newcommand{\gtwo}{\texttt{G$_2$}\xspace}
\newcommand{\spinra}{\ensuremath{\alpha_0}\xspace}
\newcommand{\spindec}{\ensuremath{\delta_0}\xspace}
\newcommand{\psid}{\ensuremath{\texttt{P}_{\mathrm{sid}}}\xspace}
\newcommand{\psyn}{\ensuremath{\texttt{P}_{\mathrm{syn}}}\xspace}
\newcommand{\ab}{\texttt{a/b}\xspace}
\newcommand{\ac}{\texttt{a/c}\xspace}
\newcommand{\sig}{\texttt{sig}\xspace}

\newcommand{\R}{\ensuremath{\mathbb{R}}\xspace}
\newcommand{\logit}{\ensuremath{\mathrm{logit}}}

\newcommand{\filtu}{\ensuremath{u}\xspace}
\newcommand{\filtg}{\ensuremath{g}\xspace}
\newcommand{\filtr}{\ensuremath{r}\xspace}
\newcommand{\filti}{\ensuremath{i}\xspace}
\newcommand{\filtz}{\ensuremath{z}\xspace}
\newcommand{\filty}{\ensuremath{y}\xspace}

\newcommand{\ftrue}{\ensuremath{f_\text{true}}\xspace}
\newcommand{\ffeat}{\ensuremath{f_\text{feat}}\xspace}
\newcommand{\PLS}{\ensuremath{P_{\textrm{LS}}}\xspace}
\newcommand{\fLS}{\ensuremath{f_{\textrm{LS}}}\xspace}

\newcommand{\numDiam}{149,000\xspace}
\newcommand{\numPer}{51,000\xspace}
\newcommand{\numSpin}{60,000\xspace}
\newcommand{\numDamit}{11,000\xspace}
\newcommand{\numDens}{500\xspace}
\newcommand{\numTI}{2,000\xspace}
\newcommand{\numVIS}{429,000\xspace}

\begin{document}

\title{\soccaname (\socca)}

\author{%
	K.~O.~Xenos\inst{\ref{i:oca},\ref{i:mauca}}  \and
	B.~Carry\inst{\ref{i:oca}}  \and
	J.~Peloton\inst{\ref{i:icjlab}}   \and
	M.~Mahlke\inst{\ref{i:ias},\ref{i:utinam}}  \and
	J.~Berthier\inst{\ref{i:lte}} \and
  P.-A.~Mattei\inst{\ref{i:inria}} 
}

\institute{
	Universit{\'e} C{\^o}te d'Azur, Observatoire de la C{\^o}te d'Azur,
	CNRS, Laboratoire Lagrange, France
	\label{i:oca}
	\and
	MAUCA - Master track in Astrophysics, Universit{\'e} C{\^o}te
	d'Azur \& Observatoire de la C{\^o}te d'Azur, Parc Valrose, 06100,
	Nice, France
	\label{i:mauca}
	\and
	Universit{\'e} Paris-Saclay, CNRS/IN2P3, IJCLab, 91405 Orsay, France
	\label{i:icjlab}
	\and
	Institut d'Astrophysique Spatiale, Université Paris-Saclay, CNRS,
	F-91405 Orsay, France
	\label{i:ias}
  \and
	Universit{\'e} Marie et Louis Pasteur, CNRS, Institut UTINAM (UMR 6213), {\'e}quipe Astro, 25000 Besan\c{c}on, France 
	\label{i:utinam}
	\and
	IMCCE, Observatoire de Paris, PSL Research University, CNRS, Sorbonne Universit{\'e}s, UPMC Univ Paris 06, Univ. Lille,
75014 Paris, France
	\label{i:lte}
  \and
  Universit{\'e} C{\^o}te d'Azur, Inria, Maasai project-team, 
  Laboratoire J.A. Dieudonn{\'e}, UMR CNRS 7351,
  France
  \label{i:inria}
}

\date{Received September 15, 1996; accepted March 16, 1997}

\abstract
{Large photometric surveys provide sparse multi-band photometry for millions of Solar 
System objects (SSOs), offering an opportunity to jointly constrain their physical and 
compositional properties. However, current phase function models do not account for rotational 
variability, limiting their ability to retrieve accurate parameters. Similarly, methods
that recover shape and rotational parameters remain both computationally and observationally expensive, making
the extraction of such properties prohibitive at scale.}
{We aim to develop a model capable of simultaneously retrieving the absolute magnitude, 
phase parameters, spin state, and shape proportions of SSOs from sparse photometric data, 
while remaining computationally efficient for large datasets.}
{We introduce the \soccaname (\socca), which extends the 
\hgg formalism by incorporating the projected surface of a rotating triaxial ellipsoid. The model 
jointly fits multi-band photometry, and includes a dedicated treatment of rotational period 
determination. }
{We implement the model on a 10-year \lsst simulation as well as on real data of asteroid (45) Eugenia for validation purposes.
\socca significantly improves the fit to photometric data, reducing the mean residuals to
half, compared to previous models. It retrieves the absolute magnitude with a scatter about three 
times smaller than existing approaches, and improves the determination of phase parameters by a similar factor. It also
recovers the sidereal rotation period, spin axis orientation and the axes ratios of the best fitting ellipsoid.
The inclusion of shape and rotation increases the number of physically meaningful solutions by $\sim$10-20\% 
per filter, leading to an overall success rate of 53\%.}
{By combining phase, shape, and rotational information in a single model, 
\socca provides a more complete physical description of SSOs from sparse photometry. 
Its performance and scalability make it well suited for current and upcoming large surveys such as 
the Zwicky Transient Facility (\ztf) and the recently started Legacy Survey of Space and Time (\lsst).}


\maketitle
%
%
\section{Introduction}

The asteroids and comets, hereafter referred to as
Solar System Objects (SSOs), are the remnants of
the planetesimals, the building blocks that accreted
to form the planets 4.6 Gy ago
\citep{2015SciA....1E0109J, 2022NatAs...6...72M}.
Their composition, studied via remote sensing observations and laboratory
analysis of meteorites,
inform us on the place and timing of their formation in the
protoplanetary disk around the young Sun
\citep{2018-ApJ-854-Scott, 2019ApJ...875...30N}.
The subsequent stages of planetary formation such as
planet migration
and
dynamical instability
\citep[e.g.,][]{2011Natur.475..206W,2020MNRAS.492L..56C}
defined how these compositions are distributed in the Solar system
\citep{2014Natur.505..629D}.

Their current distribution is, however, a degraded version of this
original, post-planetary formation, distribution.
First, gigayears of collisions have
created localized over-representation of compositions
\citep[the dynamical families,][]{1918AJ.....31..185H,2022CeMDA.134...34N}.
Second, the secular dynamical evolution
described by the Yarkovsky effect
\citep{2015-AsteroidsIV-Vokrouhlicky}
has slowly
diffused structures \citep{2001Sci...294.1693B}.
This process of collisional grinding and orbital drift is responsible
for the injection of objects in
near-Earth space: the near-Earth objects and ultimately the meteorites
\citep{1998Icar..132..378F},
our compositional references
\citep{2022Icar..38014971D}.

A deep understanding of the original distribution of planetesimals
hence requires a large corpus of SSOs with both dynamical and compositional information
together with their long-term evolution. The present study focuses on the
development of a method to jointly
determine the colors and physical properties of SSOs
to provide observational constraints on both aspects.

Multiple colors, the difference of magnitude in two photometric filters, 
over the visible and near-infrared wavelength ranges
can be used to taxonomically classify SSOs
\citep{1982Sci...216.1405G,2013Icar..226..723D, 2018A&A...617A..12P}.
Each class is based on spectral features
\citep{1984-PhD-Tholen, 2022A&A...665A..26M} and
can be used as proxy for composition thanks to
dedicated
analyses of individual SSOs and comparison with
meteorites,
returned samples,
and minerals in the lab
\citep{
  1970Sci...168.1445M,
  2006Sci...314.1711B,
  2011Sci...333.1113N,
  2015-Icarus-252-Cloutis,
  2023Sci...379.8671N,
  2024M&PS...59.2453L,
  2026A&A...705A.121M}.

The long-term evolution, due to semi-major drift by
the non-gravitational Yarkovsky effect, is directly affected by the 
physical parameters of the SSO: diameter, albedo, obliquity, rotation period,
density, thermal inertia \citep{2015-AsteroidsIV-Vokrouhlicky}.
It has been shown that Yarkovsky, and
the associated torque YORP (\textit{ibid}) which changes the spin,
are not constant over long timescales but have a
stochastic evolution owing to small changes on the surface of
these objects \cite[craters, landslides, see][]{2009Icar..202..502S,
  2015Icar..247..191B, 2024A&A...682A.130Z}.
We lack a large sample of SSOs with measured physical
parameters to fully assert the timescale and amplitude of
this stochastic evolution, despite
recent progresses \citep{2025NatAs...9..493Z}.

The sample of SSO with
taxonomic classification
has been limited to small number statistics for decades.
Most determinations were obtained by targeted surveys, aiming at
obtaining the spectral reflectance or the colors of the objects
\citep{1970Sci...168.1445M, 1985Icar...61..443M, 1996P&SS...44..555B,
	2008ApJ...682L..57M}.
This corpus has dramatically increased over the last decade, mainly
thanks to large sky surveys that provided near-simulatenous multi-filter
observations \citep[such as the SDSS, SkyMapper, VISTA, see][]{
	2018PASA...35...10W,
	2013Msngr.154...35M,
	2010A&A...510A..43C,
	2013Icar..226..723D,
	2018A&A...617A..12P,
	2021A&A...652A..59S,
	2022A&A...658A.109S}
or spectra \citep[Gaia,][]{2016A&A...595A...1G,
	2023A&A...674A..35G,2023A&A...674A..12T}.
The current sample\footnote{%
All numbers are extracted from the \ssodnet service \citep{2023A&A...671A.151B}, 
arguably the largest collection of properties of SSOs:
\url{https://ssp.imcce.fr/webservices/ssodnet/}

The spectra themselves can be accessed through
\texttt{classy}:\\
\url{https://github.com/maxmahlke/classy}
}
of visible spectra contains over 60,000 asteroids,
visible colors are available for \numVIS, and
near-infrared colors for about 35,000.

The determination of physical parameters has followed a parallel evolution.
From the first historical
diameter and albedo determinations
\citep[e.g.,][]{1974ApJ...194..203M, 1979AJ.....84..259W},
IRAS \citep{1984ApJ...278L...1N} was the first survey to provide 2000 asteroids
\citep{2000Icar..146..161S}, and the all-sky mid-infrared
space missions WISE \citep{2010AJ....140.1868W} and
AKARI \citep{2007PASJ...59S.369M} brought the sample to
\numDiam \citep{2011PASJ...63.1117U, 2013PASJ...65...34H,
	2011ApJ...743..156M, 2011ApJ...742...40G, 2011ApJ...741...68M,
	2012ApJ...747...49B}.
Rotation periods were traditionnally studied by acquiring
time-serie photometry, dense in time \citep[generally refered to as
	light curves,
	see,][]{1906ApJ....24....1R, 1975A&A....44...81S,
	1987Icar...72..325F, 2012MPBu...39..171P},
a highly telescope-time-consuming
process.
Data mining of wide-field photometry surveys built for
transients sky \citep[such as the PTF,][]{2009PASP..121.1395L} or
exoplanet detection by the transit method \citep[e.g., Kepler/K2, TESS,][]{
	2010Sci...327..977B,
	2014PASP..126..398H,
	2015JATIS...1a4003R}
have recently produced many rotation periods
\citep{2015AJ....150...75W,
	2015ApJS..219...27C,
	2020ApJS..247...26P,
	2023ApJS..264...18K,
	2025A&A...693A..66V}.
Yet, the total number remains limited to
\numPer, much smaller that the known population of 1.5 million SSOs.

The remaining properties are the least constrained:
spin orientation (spin coordinate and obliquity, about \numSpin SSOs),
density (\numDens) and thermal inertia (\numTI).
Density requires mass estimate, which is arguably the most difficult property
to measure for SSOs \citep{2012PSS...73...98C, 2015-AsteroidsIV-Scheeres}.
The large collection of mid-infrared data from IRAS, AKARI and WISE,
and the upcoming NASA NEO Surveyor 
and ESA NEOMIR missions \citep{2023PSJ.....4..224M, 2024SPIE13092E..2HC},
could lead to
the determination of thermal inertia for numerous SSOs
through thermophysical modeling \citep{1996A&A...310.1011L,
	2021AJ....162...40J}.
The limiting factor is, however,
the availability of the spin properties (and shape).

Light curves obtained under different Sun-Target-Observer geometries
(i.e., generally obtained over multiple years) have been used for
50 years to determine the spin and
simple triaxial-ellipsoid shape of asteroids
\citep[e.g.,][]{1977A&AS...30..121S, 1984Icar...57..443O}.
A major improvement occurred at the turn of the XXth century with what
is now called the
standard light curve inversion \citep[LCI,][]{2001Icar..153...24K,
	2001Icar..153...37K},
providing a robust method for spin and convex shape determination
\citep[other successful approaches are also in use, such as][]{
	2018MNRAS.473.5050B,  2019A&A...631A..67C,  2020A&A...642A.138M}.
The LCI approach is also applicable to sparse photometry
\citep[that is photometry acquired with a time sampling typically larger
	than the rotation period of the target,][]{2004A&A...422L..39K}
partly overcoming the limitation imposed by the telescope-time
comsuming aspect of
light curves.
In twenty years, the application of the LCI to Lowell, WISE, Atlas, Gaia
photometry brought the number of spin and shape determination to
almost \numDamit asteroids
\citep[e.g.,][]{2016A&A...587A..48D, 2018A&A...617A..57D, 2020A&A...643A..59D}.
Nevertheless, the fraction of successful fits compared to the sample of
SSOs (hereafter the success rate) remains limited and time consuming to obtain.
\citep[see][]{2015A&C....13...80D, 2024A&A...687A.277C, 2023A&A...675A..24D}

The interest of using large sky surveys for studying
the composition and physical properties of SSOs is clear
owing to the amount of observations they produce.
This is particularly true for the LSST survey of the
Vera C. Observatory \citep{2019ApJ...873..111I} which is expected to
return around 400 observations in six filters 
\citep[\filtu, \filtg, \filtr, \filti, \filtz, \filty,][]{2024SPIE13096E..1SR}
for about five million SSOs \citep{2025AJ....170...99K}.
The challenge remains in identifying SSOs in these surveys
(source of potential outliers from mis-associations), and
handling data not solely optimized for SSO science. This mostly applies to
cadence, i.e., the time interval between frames, that strongly select
the SSOs based on their apparent motion, and may preclude the direct
computation of colors if the acquisition of different filters is too
separated in time \citep[see the discussions
in][]{2016A&A...591A.115P, 2018A&A...609A.113C}.

Here, we build upon the recent \shgg model
\citep{2024A&A...687A..38C} to combine two approaches, phase function and
spin/shape modeling, into a general approach:
\soccaname (\socca).
It provides the
absolute magnitudes in multiple filters, hence the colors,
together with the spin and shape of SSOs, from sparse, multi-filter
photometry such as provided by large sky surveys.

We present the \socca model in \Cref{sec:model} and the determination
of its initial parameters, crucial for a successful fit, in 
\Cref{sec:init}. We go into more detail for the initialization of the rotational period in \Cref{sec:init:period} and
we then validate our approach in 
\Cref{sec:valid} based on simulated observations of SSOs on real data.
In \Cref{sec:validate_eugenia} we present an evaluation of the algorithm's performance on real data of the asteroid (45) Eugenia.
Finally in \Cref{sec:disc} we discuss certain improvements and extensions to the model and in \Cref{sec:conclusions} we summarize the
work presented and lay out some conclusions.

\section{The \socca model\label{sec:model}}

The apparent magnitude $m$ of a SSO in any filter
can be expressed as its
intrinsic absolute magnitude $H$ (in the same filter) modulated
by the changing geometry of observations:

\begin{equation}
  \label{eq:Hfgs}
	m = H + \funcf + \funcg + \funcss
\end{equation}

\noindent where the function
\funcf accounts for the
changing Sun-target and target-observer distances;
the function \funcg describes the evolution of brightness with the
phase angle ($\gamma$, the Sun-target-observer angle); and
the function \funcss the ever-changing apparent shape of the target.

Broadly, Eq. \ref{eq:Hfgs} is referred to as the phase function.
The first version was written by \citet{1989-AsteroidsII-Bowell}.
Since 2012, the IAU recommendation
is to use the following formalism by \citet{2010Icar..209..542M}:

\begin{eqnarray}
	\funcf &=& 5 \log_{10}(r\Delta) \\
	\funcg &=& - 2.5\log_{10}\left[ \gone \phi_1(\gamma) \right.\nonumber \\
		&& \hspace{3.1em} +   \gtwo \phi_2(\gamma)\nonumber \\
		&& \hspace{3.1em}\left. + ( 1 - \gone -\gtwo) \phi_3(\gamma) \right]
\end{eqnarray}

\noindent We adopt the updated constraints on \gone\ and \gtwo\ proposed by \cite{OSZKIEWICZ2026116886}:

\begin{subequations}
  \label{eq:gg}
  \begin{align}
    \gone & \geq -0.429  \label{eq:cond_G1}           \\
    \gtwo & \leq 1.429 \label{eq:cond_G2}           \\
    \gtwo & \geq -0.4 \, \gone \label{eq:cond_GG3} \\
    \gtwo & \geq -3.9038 \, \gone - 0.2445 \label{eq:cond_GG1}        \\
    \gtwo & \leq -0.9635 \, \gone + 1.0157 \label{eq:cond_GG2}      
  \end{align}
\end{subequations}

The \funcss function was only introduced recently by
\citet{2024A&A...687A..38C} to solve a common issue of phase function: the
variation of absolute magnitude from apparition to apparition
\citep[e.g.,][]{2021Icar..35414094M, 2025Icar..43616577C}.
This variation is due to the slowly-changing geometry
(i.e., the angle between the target spin axis and the viewing direction,
called the aspect angle $\Lambda$)
under which the shape of the SSO is seen
\citep{2022MNRAS.513.3076J}.
Splitting observations by apparitions only provides a partial solution:
the aspect angle evolves within an apparition
and the \ggparams parameters are
thus biased as the \funcg function attempts to fit a signal which is
a combination of phase- and shape-related variability
\citep[see Fig.~2 in][]{2024A&A...687A..38C}.

The introduction of this shape-related component to the phase function
practically solves the apparition-to-apparition effect:
it increases the success rate of the model fit,
it improves residuals between observations and model,
and the \ggparams parameters are more clustered per taxonomic class
\citep[see][]{2024A&A...687A..38C}.
The recently introduced \funcss function described
the geometry of a simple oblate spheroid
(of diameters $a = b \ge c$),
i.e., it was explicitly independent of time
(that is $\funcss = \funcs$, see \Cref{app:socca}).
However, not all SSOs are simple spheroids and many are elongated
\citep{2018ApJS..239....4T}. This leads to a non-modeled high-frequency
component in the residuals, linked with the rotation of the targets.
Furthermore, the derived spin coordinates
(\spinra, \spindec) are ambiguous
\citep[as intuited by][]{1906ApJ....24....1R},
with a mirror solution at
(\spinra+180\degr, $-\spindec$) fitting equally-well the observations.

We thus generalize here the shape function \funcss to account for the rotation of
a triaxial ellipsoid.
The proposed \socca approach is conceptually close to the ellipsoid fit
by \citet{2019A&A...631A..67C, 2024A&A...687A.277C}, but uses
the IAU-approved phase function \funcg instead of
the more restrictive linear slope.
This model is also resemblant to the
LCI approach
\citep[in which the phase function is described by
  a linear-exponential function,][]{2001Icar..153...37K}, but strives
1) to fit the apparent magnitudes (LCI handles relative fluxes) to derive
the colors of the target from the absolute magnitudes obtained in different
filters, and
2) to maintain a limited number of free parameters
\citep[ellipsoid case, which also provides a significantly faster
  computation,][]{2016A&A...587A..48D}.

We model the shape of an SSO by an
ellipsoid of semi-axes $a > b > c$,
in a target-centric reference frame,
which axes ($x$, $y$, $z$) are
aligned with the longest, middle, and shortest axes. 
The ellipsoid rotates with a sidereal period \psid along the $z$ axis, whose coordinates
are (\spinra, \spindec) in the equatorial reference frame (taken at J2000
epoch, i.e., EQJ2000).
Following \citet{1984Icar...57..443O}, we compute the brightness of the ellipsoid
as its projected area on the plane of the sky, accounting for both the 
limb and the terminator (the complete description is provided in \Cref{app:socca}).
The definition of the absolute magnitude \Hmag must therefore be updated and 
it is the magnitude
of the object at 1\,au from both the Sun and the observer,
with a 0\degr phase angle,
\textit{seen from its equatorial plane
and its prime meridian}.

\begin{figure}[t]
  \input{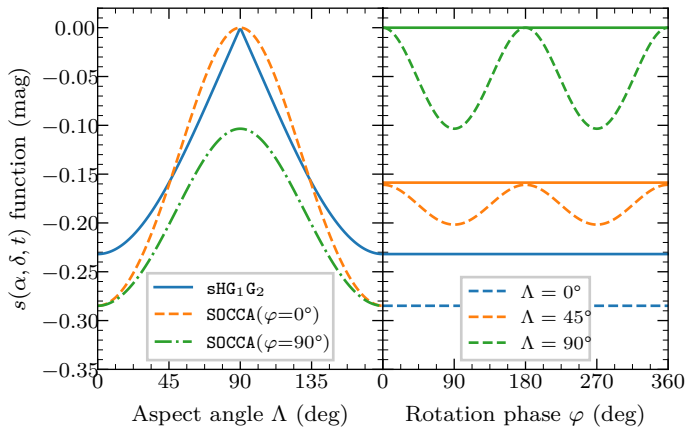}
  \caption{Comparison of the behavior of
    shape function \funcss 
    of \shgg (solid lines) and
    \socca (dashed lines)
    against aspect angle ($\Lambda$, left) and
    rotation phase ($\varphi$, right).
    \label{fig:model}
  }
\end{figure}

We compare in \cref{fig:model} the
\funcs \citep{2024A&A...687A..38C} and
the proposed \funcss functions
defining the \shgg and \socca models. The difference is most notable
against the rotation phase: the \shgg model is invariant while
\socca presents a double-peaked light curve built in \funcss as
a rotating ellipsoid.
The two models also differ against the aspect angle.
The \funcss is
smoother than the ad hoc function of \shgg
and presents a larger amplitude than \funcs.

In summary, \socca captures both the short- and long-term photometry of
SSOs with a frugal approach, i.e., with a minimalistic number of free parameters:
\begin{itemize}
  \item[$\bullet$] Three (\Hmag, \gone, \gtwo)
    for the behavior against phase angle,
  \item[$\bullet$] Two (\spinra, \spindec) for the 
    orientation of the spin axis,
  \item[$\bullet$] Two (\psid, $W_0$) for the rotation 
    around the spin axis,
  \item[$\bullet$] Two ($\ab$, $\ac$) representing the shape of the 
    ellipsoid.
\end{itemize}

Among these, the latter six parameters describing the geometry are 
wavelength independent.
The former three parameters can, however, be different for each observed
filter:
the colors of SSOs result from different \Hmag in different filters
and it has been shown that \ggparams are wavelength-dependent
\citep[and responsible for the so-called phase reddening, e.g.,][]{
2012Icar..220...36S, 2021Icar..35414094M}.
Hence the total number of free parameters is
3\nfilter+6 for observations acquired in \nfilter 
different filters.
It represents five more than \hgg \citep{2010Icar..209..542M},
but only three more than \shgg \citep{2024A&A...687A..38C},
to account for the full geometry.
Even with the six filters of LSST (i.e., $\nfilter=6$), the 
model is over-constrained (i.e. there are many more data points than free parameters) for large numbers of observations and
the parameters can be determined by $\chi^2$ minimization, as long as 
there is adequate phase and aspect (i.e. line-of-sight - spin axis) angle coverage in the photometric data.
That is required in order to constrain the \gone-\gtwo and \ab-\ac parameters respectively.

\section{Initializing the model\label{sec:init}}

\begin{figure*}[t]
  \input{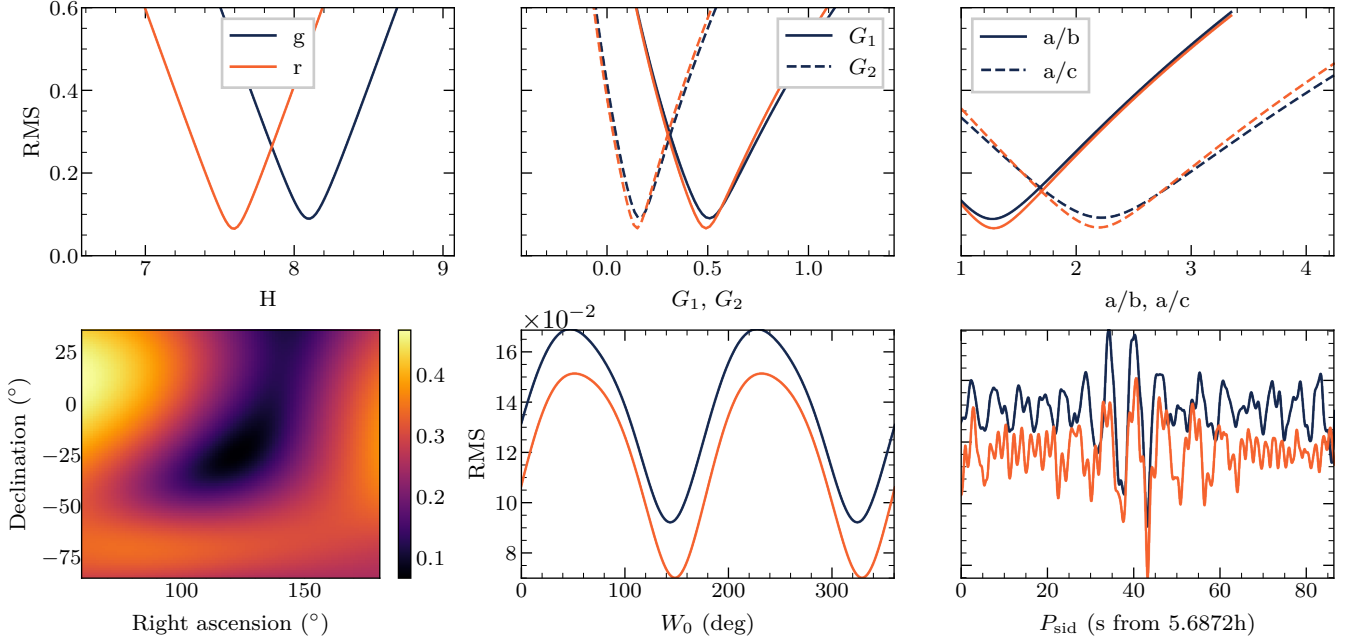}
  \caption{Dependence of \socca on its parameters
    illustrated with
    (45) Eugenia.
    All but the sidereal periods have a smooth convergence
    to the best-fitting values.
    \label{fig:converge}
  }
\end{figure*}

While \socca strives to minimize the number of free parameters,
the parameter space still represents a large volume to explore and 
contains many local minima.
The convergence of the phase-related function \funcg is 
expected to be smooth (\Cref{fig:converge}) provided
observations at low phase angle are
secured \citep[below $\approx$5\degr, see][for a discussion on the need for 
near-opposition observations]{2021Icar..35414094M}.
The shape-related function \funcss is more likely to 
present numerous local minima, for the 
spin coordinates (\spinra, \spindec) and even more for
the sidereal rotation 
period \psid \citep[a well-documented aspect of shape
inversion, see][for a review]{2015-AsteroidsIV-Durech}.

We tackle this issue here with the objective of remaining frugal.
The number of SSOs will strongly increase with LSST
\citep{2025AJ....170...99K} and the number of observations
even more \citep{lsst2009}. To fit all these observations
while minimizing the needed resources and their impact, 
we focus on limiting computations rather than expensive grid searches.
We thus use a combination of \shgg fit and
frequency analysis 
to 
estimate the optimum initial conditions to determine the
parameters of \socca through traditional gradient-descent
least-square minimization.
We describe in this section the initialization of most parameters and
discuss the more complex rotation period next.

\subsection{Phase and shape\label{ssec:init:shape}}

While the definition of \Hmag differs between
\shgg and \socca, it only acts as a small global photometric offset, easily 
captured by the model (\Cref{fig:converge}). The phase function \funcg
parameters \ggparams are strictly similar on the other hand.
For these reasons, we directly initiate \socca inversion with the \Hmag, \gone, and
\gtwo obtained from \shgg modeling.

The description of the shape between the two model improves from oblate spheroid
to triaxial ellipsoid. The residuals of the \shgg modeling are thus due to the
non-modeled short-term variability, linked with \ab, which can in turn be
estimated from the peak-to-peak amplitude $A$ of the residuals.
This together with the
relation between the oblateness $R$ and the axes ratios
\citep[\Cref{eq:R},][]{2024A&A...687A..38C}
give:
\begin{eqnarray}
  \ab &=& 10^{0.4 A} \\
  \ac &=& \left(\ab +1\right) \left/ 2R \right.
\end{eqnarray}

\subsection{Spin axis orientation\label{ssec:init:spin}}

The \shgg fit provides a spin solution (\spinra, \spindec) and its
mirror solution (\spinra+180\degr, -\spindec).
Rather than initializing \socca at these two coordinates only,
we perform an additional safety check.
We compute the root mean square (RMS) of \shgg residuals on a grid
of spin-axis coordinates spanning the entire celestial sphere:
\spinra $\in$ [0\degr, $360\degr[$  
and
\spindec $\in$ [-90\degr, $90\degr]$ with steps of
10\degr and 5\degr, respectively.  
We only do forward computation of \shgg, keeping all parameters but the
spin coordinates fixed. It is therefore a computationally light and 
fast grid.

We smooth the resulting RMS map with a Gaussian kernel density estimate
of width $4\degr$ to suppress high-frequency noise. This scale is comparable
to the typical uncertainty on $(\alpha, \delta)$, and the exact choice has
little impact on the result.
Local minima are then identified on the interpolated map
by running a ${7.5}\degr \times {3.75}\degr$ two-dimensional window minimum filter.
The resulting list of (\spinra, \spindec) solutions are each tested as initial
points for the \socca fit.

\section{Rotational period \label{sec:init:period}}
\subsection{Finding the synodic period \label{ssec:syn_period}}

Photometric data from sky surveys typically span many years:
8\,y for ZTF \citep{2019PASP..131a8002B} and 
9\,y for ATLAS \citep{2018PASP..130f4505T} that
we will use later to illustrate \socca on real data
(\Cref{sec:validate_eugenia})
and 
10\,y foreseen for the LSST \citep{2019ApJ...873..111I}.
The rotation period of SSOs are much shorter, from about 2\,h
to a few days: half of known periods are shorter than 12\,h
\citep{2023A&A...671A.151B}
but there is a bias against observing slow rotators,
see \citet{2018A&A...610A...7M}.
Hence, any deviation from the true sidereal period \psid
in the initialization of \socca fit
would lead to a significant dephasing of the viewing geometry
(the subobserver longitude $\varphi$, \Cref{app:socca})
over the observation epochs.
Owing to the forest of local minima in the sidereal
period \psid space
(\Cref{fig:converge}),
a robust method for its initialization is required.

As stated in \Cref{ssec:init:shape}, the residuals of 
\shgg are due to the non-modeled rotation.
We thus search for periodicity in these residuals, using
Lomb-Scargle frequency analysis 
\citep{1976Ap&SS..39..447L,1982ApJ...263..835S}.
It is non-trivial and caution must be applied.
First, previous works have shown the strong effect of the observational cadence
on the determination of rotation period \citep{2022FrASS...909771D}.
Second, the determined period only corresponds to the
synodic rotation period (\psyn) of the SSO
as observed from Earth and not its
sidereal rotation period (\psid).

We use the \nifty implementation of Lomb-Scargle
\citep{Garrison_2024}, providing a performance boost of
almost two orders of magnitude compared to
the traditional algorithm \citep{1989ApJ...338..277P}.
Because observations are
unevenly-sampled in time,
some over-sampling of the expected width of each frequency
peak is required \citep{2018ApJS..236...16V}.
For each SSO, the pseudo-Nyquist frequency
($\delta f$) is evaluated
\citep[with five sampling points per
  peak,][]{1996ApJ...460L.107S, 2015ApJ...812...18V}
as
\begin{equation}
  \delta f = 5 / \Delta T
\end{equation}

\noindent with $\Delta T$ the baseline.
The frequency limits used for the periodogram are 
$0.12$ and $2.4 \times 10^5$ hours,
corresponding to a sensitivity to rotation periods between 
$0.24$ and $1.2 \times 10^5$ hours
(choice based on the distribution of known asteroid rotational periods). 
However, because the parameter space is large and the computation becomes costly, we adopt a two-step strategy.
First, a period search window between $1.2$ and $2.4 \times 10^5$ hours
is used and the Lomb-Scargle periodogram is computed using an
increasing number of sine and cosine terms $k$, from 1 to 4,
similarly to the approach proposed by \citet{2025A&A...693A..66V}.
The F-statistic is then computed for each consecutive pair of periodograms in the following
manner:

\begin{equation}\label{eq:f-test}
  F = \frac{\nobs - \mathrm{dof}_{k+1}}{\mathrm{dof}_{k+1} - \mathrm{dof}_{k}}
  \times \left(\frac{\mathrm{rms}_{k}^2}{\mathrm{rms}_{k+1}^2} - 1\right)
\end{equation}

\noindent where \nobs is the number of data points used in the fit, 
$\mathrm{dof}$ the degrees of freedom, and 
$\mathrm{rms}$ the root-mean-square residuals.
The indices $k$ and $k+1$ indicate the number of terms of the two compared periodograms.
Finally, the 
degrees of freedom are computed
as $2k+1+3\nfilter$, with \nfilter the number of filters used in the photometric data.

\indent To assess whether the increase in model complexity
($k \rightarrow k+1$) is statistically
justified, we compare the measured $F$ value to a critical threshold 
$\alpha_F$ derived from the $F$-distribution. 
This critical value corresponds to the $99^{\mathrm{th}}$ percentile of the distribution.
A more complex model $k+1$ is retained only if $F > \alpha_F$.
If this condition is not satisfied, the improvement in the fit is not considered significant at the $1\%$ level, 
and the simpler model $k$ is preferred, according to the Occam's razor rule.
If more than $k=4$ terms provide a statistically significant improvement to the fit,
we consider that no period in the range $1.2 - 2.4 \times 10^5$ hours
describes the variability in the data.
In the aforementioned case, we repeat this $F$-test procedure on shorter periods in the
$0.12 - 1.2$ hours range. The procedure is repeated in this two-step manner
because the period search in the very high frequency space takes significantly longer.

\subsection{Identifying true periods from aliases}\label{ssec:alias_period}

\begin{figure}[t]
  \input{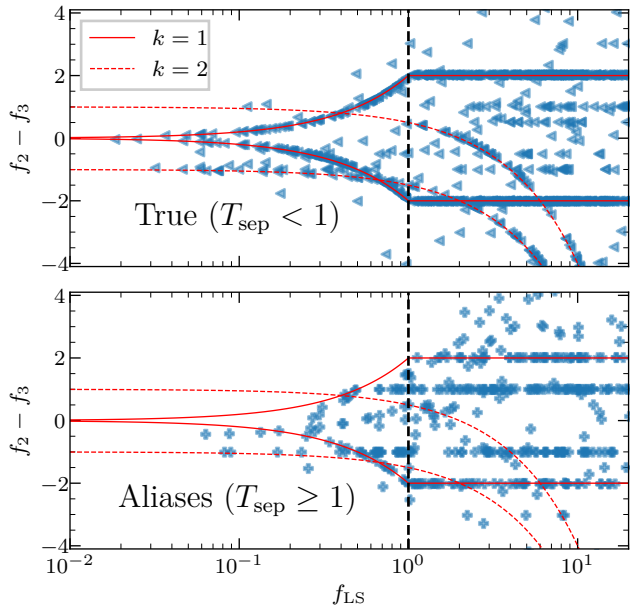}
  \caption{Difference between the second and third highest periodogram peaks,
$f_2-f_3$, as a function of the observed frequency $f_\text{LS}$. The red curves show the
analytical relations expected for true frequency evaluations from
\Cref{eq:trumpet}: solid lines correspond to the $k=1$ relation,
while dashed lines correspond to $k=2$.
\textbf{Top}: solutions identified as true periods. The majority of the points follow the relations of \Cref{eq:trumpet}.
\textbf{Bottom}: solutions identified as aliases. Most of the frequency differences
$(f_2-f_3)$ do not follow the predicted relations.
The vertical dashed line indicates the feature frequency, i.e., the 24\,h
observational cadence.
}
\label{fig:trumpet}
\end{figure}

The periodogram typically exhibits a dominant peak, which may correspond
to the true synodic period, one of its aliases, a period linked with the cadence of observations,
or simply a noise-driven outlier. Given the extreme sensitivity of the photometric model
to the initial period evaluation (\Autoref{fig:converge}), a method is required
to identify in which category the period evaluation falls in.

We implement a test based on bootstrap
inspired by a previous analysis of ATLAS sparse photometry \citep{2022FrASS...909771D}.
We hold fixed the temporal sampling 
while each bootstrap realization is generated by randomly drawing
(with replacement) a set of photometric measurements from the 
original dataset.
For each resampled dataset, the period search is repeated on the corresponding residuals.

A key distinction from earlier studies is that the periodograms are computed
using the residuals of the \shgg photometric model rather than 
the raw light curves.
This choice ensures that 
the variability of the signal due to the varying aspect angles in different
apparitions are removed by the model, leaving only signatures 
introduced by the intrinsic rotation and the observational cadence. Therefore, 
a simple Lomb-Scargle periodogram is enough to find the period,
without performing a full modelisation of the SSO,
saving significant computation time.
We generate a total of 25 bootstrap realizations and record for each
the period for the highest periodogram peak. 
For the i-th bootstrap sample, we compute the relative deviation

\begin{equation}\label{eq:bootstrap}
  \Delta P = \frac{|P_{\text{BS},i} - \PLS|}{\PLS}
\end{equation}

\noindent where $\PLS$ denotes the period associated with
the dominant peak in the periodogram of the original residuals, and $P_{\text{BS},i}$
is the corresponding bootstrap estimate. The bootstrap score 
$N_{BS}$ is then defined as the number of bootstrap samples satisfying 
$\Delta P \le 0.01$ providing a measure of the period's evaluation 
stability (chosen here at 1\%).

The observing window and its harmonics, which are typically observed
at 24 hours and its perfect divisors and multiples,
are responsible for creating strong aliases in the periodogram.
The observing window is a secondary periodic 
signal in the time series of the residuals due to
the observation cadence of the telescope, which is convolved
with the primary asteroid spin period signal, giving rise to the
multiple peaks around the true periodic signal. Nonetheless, the
strongest peak in the periodogram is not necessarily due to the
asteroid's spin, but it can be one of the features described above.
In general, the peak of the periodogram is given by:
\begin{equation}\label{eq:aliasfunc}
  \fLS = \frac{1}{\PLS} = \frac{i+1}{\psyn} \pm \frac{j}{24 \text{ hours}}, \quad i,j > 0
\end{equation}

\indent To discriminate between true rotational frequencies and aliases produced by the observing window, 
we adopt a flagging scheme based on the relative location and ordering of the strongest peaks in the periodogram.
Let \fLS $=1/P_\text{LS}$ be the frequency corresponding to the highest peak of the Lomb-Scargle periodogram, 
and let $f_{\mathrm{feat}}$ denote the characteristic frequency of the observing window.  
In practice, $f_{\mathrm{feat}}$ is typically associated with the 24-hour cadence.
We further define the peak distance
\begin{equation}
\Delta f_1 = f_{2} - f_{3}
\end{equation}

\noindent
where $f_{2}$ and $f_{3}$ are the frequencies of the second and third highest peaks in the periodogram, respectively.  
The sign of $\Delta f_1$ provides information on the asymmetry of the peak
structure around the dominant feature (\Cref{fig:trumpet}).
The alias-flagging procedure depends on the number of terms $k$ selected for the periodogram by the F--test criterion
(\Cref{eq:f-test}).

In the case where $k=1$ we rely on the following assumptions.  
First, the highest peak in the periodogram corresponds to the true rotational frequency of the asteroid, 
while aliases induced by the observing window appear as secondary peaks on either side of this frequency in frequency space.  
Second, we assume that no additional significant peaks are produced by harmonics of the rotational signal itself, 
such that the secondary structure of the periodogram is dominated by window-induced aliases.

In the case where $k=2$ we again assume that the highest peak in the periodogram corresponds
to the true rotational frequency of the asteroid.
The highest peak corresponds to the first harmonic and the second highest
peak is the fundamental, at $\fLS = 2 \ftrue$.
The third highest peak can then be on either side of the highest peak of the periodogram.
These assumptions are driven by the visual inspection of numerous periodograms produced using our simulated dataset.

Using these assumptions, one can find from
\Cref{eq:aliasfunc}
the following diagnostic quantity $T_{k}$ as
\begin{equation}
  \label{eq:trumpet}
T_k =
\begin{cases}
+2 \ffeat, & k = 1, \;\; \Delta f_1 > 0 \;\; \text{and} \;\; \fLS > \ffeat, \\[2pt]
-2 \fLS, & k = 1, \;\; \Delta f_1 < 0 \;\; \text{and} \;\; \fLS < \ffeat, \\[2pt]
+2 \fLS, & k = 1, \;\; \Delta f_1 > 0 \;\; \text{and} \;\; \fLS < \ffeat, \\[2pt]
-2 \ffeat, & k = 1, \;\; \Delta f_1 < 0 \;\; \text{and} \;\; \fLS > \ffeat, \\[4pt]
-0.5 \fLS - \ffeat, & k = 2, \;\; f_2 > \fLS, \\[2pt]
-0.5 \fLS + \ffeat, & k = 2, \;\; f_2 < \fLS \\[2pt]
\end{cases}
\end{equation}

Wether a peak is an alias or not is then quantified by the distance between the measured peak 
separation and the expected peak separation under the assumption that the peak corresponds to the true signal:
\begin{equation}
T_\text{sep} = 100 \, \left| \Delta f_1 - T \right|
\end{equation}

If $T_\text{sep} < 1$, the periodogram peak is classified as a true rotational period.  
Otherwise, the peak is flagged as an alias induced by the observing window.
The success rate of the flagging scheme is listed in table \ref{tab:purity-completeness-periods}.

\rev{We reiterate that the alias flagging method is based on the assumption of a dominant 24 hour observational 
cadence, which is representative of most ground based surveys. Nevertheless, adopting a different cadence would 
primarily modify the method quantitatively rather than conceptually, provided that equation \ref{eq:aliasfunc} remains valid.
This condition is satisfied for the vast majority of period analyses performed with the Lomb-Scargle periodogram, meaning that 
the approach can in principle be adapted to other surveys, including combined datasets, by recalibrating the corresponding $T_k$ 
values of equation \ref{eq:trumpet} for the relevant cadence structure.} 

\subsection{From synodic to sidereal period} \label{ssec:sid_period}

The Lomb-Scargle periodogram provides an average synodic period \psyn.
The relevant quantity
required to tie all observations together
is the sidereal period \psid, defined in an inertial frame.
The two periods can differ by tens of seconds for a typical period of a few hours.
Their difference is a multivariate quantity which depends on the semi-major axis of the asteroid,
its spin axis orientation and its sidereal period itself. 
Considering that the subobserver longitude between two observations $A$
and $B$ separated
by exactly one synodic period (such as $t_B = t_A + \psyn$) differ by $2\pi$
by definition, one finds 

\begin{equation}
  \label{eq:synvssid}
  \psid = 2 \pi \psyn \left/ (W_A-W_B)\right.
\end{equation}

\noindent where $W_i$ are functions of the coordinates of the target and its spin axis
(\Cref{eq:sep_long}).
The difference between the synodic and sidereal periods is therefore not fixed,
but depends on 
the period itself and the apparent motion of the target.
Those located closer to the observer exhibit a larger apparent motion on the sky over
one synodic period, hence 
a larger difference between \psyn and \psid.

\begin{figure*}[t]
  \centering
  \input{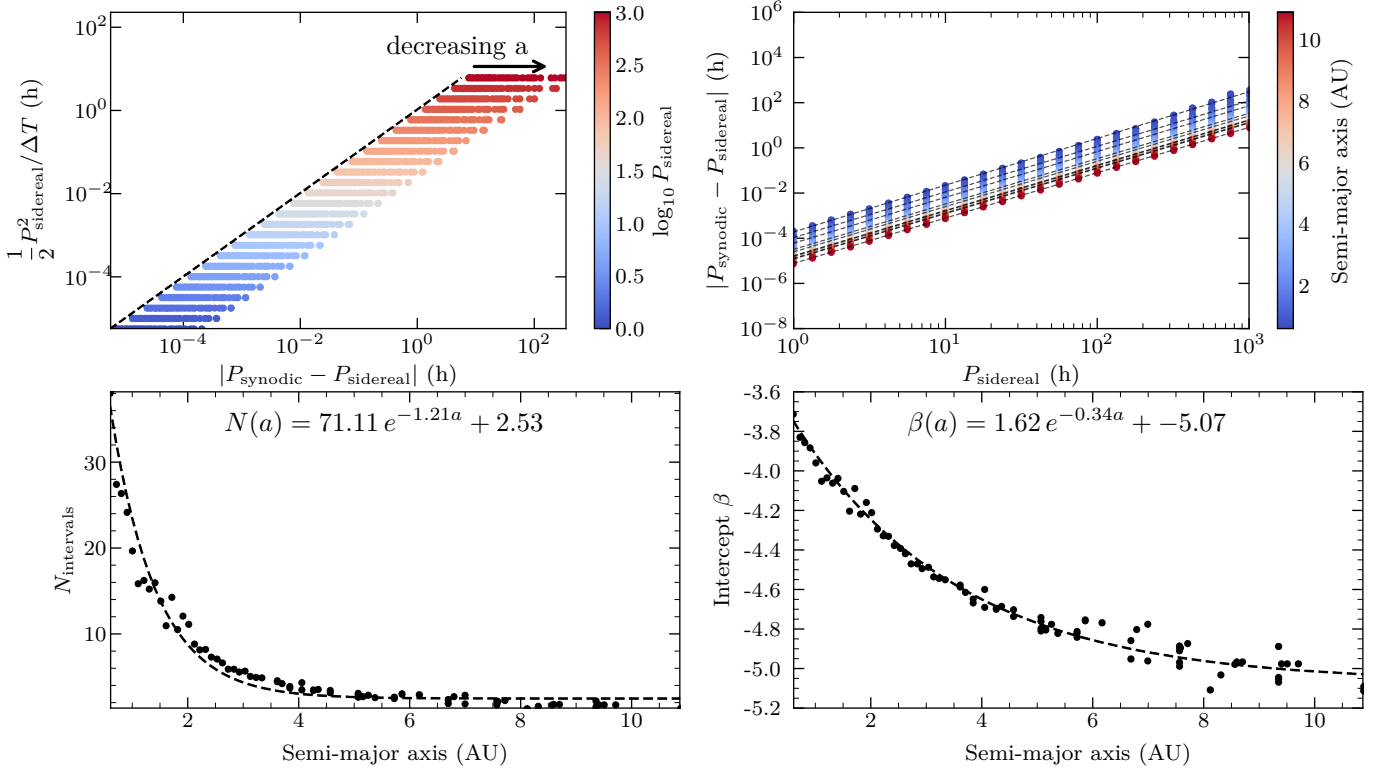}
  \caption{
Procedure for defining the synodic-sidereal period window and the number
  of resolvable period intervals as a function of semi-major axis.
\textbf{Top left:} For each semi-major-axis slice, the maximum synodic-sidereal period offset
$|P_{\mathrm{syn}} - P_{\mathrm{sid}}|$ is compared to the Lomb--Scargle periodogram resolution limit,
$\tfrac{1}{2} P_{\mathrm{sid}}^{2} / \Delta T$, where $\Delta T$ is the 10-year observational baseline.
Points are color-coded by \psid
and the black line marks the 1:1 line.
\textbf{Top right:} Maximum resolvable synodic-sidereal offset as a function of sidereal period for each semi-major-axis slice.
Dashed lines show power-law fits of the form
$\log_{10}|P_{\mathrm{syn}} - P_{\mathrm{sid}}| = 2 \log_{10} P_{\mathrm{sid}} + \beta$,
with color indicating semi-major axis.
\textbf{Bottom left:} The number of resolvable period intervals,
$N_{\mathrm{intervals}} = |P_{\mathrm{syn}} - P_{\mathrm{sid}}| / (\tfrac{1}{2} P_{\mathrm{sid}}^{2} / \Delta T)$,
as a function of semi-major axis, with an exponential approximation.
\textbf{Bottom right:} Fitted intercept $\beta$ as a function of semi-major axis, also modeled by an exponential decay.
  \label{fig:window}
}
\end{figure*}

We explore this difference 
by randomly selecting 100 SSOs with semi-major axes 
evenly space between 0.5 and 10 au.  
For each, we use the 
\miriade ephemeris
service\footnote{\url{https://ssp.imcce.fr/webservices/miriade/}}
\citep{2008LPICo1405.8374B} to compute their apparent positions on sky every two days for
two full orbital periods.
For each trial sidereal period \psid,
we compute the subobserver longitudes $\varphi_i$ 
at each propagated epoch and
derive the corresponding synodic periods (\Cref{eq:synvssid}).

The absolute difference between \psyn and \psid
at each epoch quantifies the dephasing caused by the
combined motion of the SSO and the observer.
To capture the effect of varying spin properties,
we repeat this calculation over a grid of 
\spinra $\in$ [0\degr, $360\degr[$ with steps of 10\degr,
\spindec $\in$ [-90\degr, $90\degr]$ every 5\degr,
\psid $\in$ [1\,h,  $10^3$\,h] with 25 logarithmics-spaced steps.
For each of the 1250 combinations of
semi-major axis, sidereal period and spin-axis coordinates, 
we record the median of the absolute deviation between synodic and sidereal period
(for all epochs) as shown in the top left panel of figure \ref{fig:window}.

We then compute the maximum period deviation
($\Delta P_{\max}$) for each semi-major axis (top left panel of figure \ref{fig:window}).
We use this to determine
the sample of sidereal periods to be tested from the synodic period
found by Lomb-Scargle analysis.
Specifically, for a survey of duration $T_\mathrm{survey}$,
the uncertainty in resolving a period \psid can be approximated as
\begin{equation}
  \Delta \PLS \sim \frac{1}{2} \frac{\psid^2}{T_\mathrm{survey}}
\end{equation}

which represents the minimal detectable difference between aliases
\citep{2001Icar..153...24K}.
The number of sample is then given by the 
ratio of
the maximum period deviation and the period resolution
($\Delta P_{\max} / \Delta \PLS$).

Furthermore, 
the magnitude of this synodic-sidereal difference increases with the sidereal period itself
(\Cref{eq:synvssid}). 
Slower rotators accumulate a larger phase offset over the same observing baseline,
amplifying the discrepancy between the synodic and sidereal solutions. 
Both effects are reproduced by our simulations, which show that the maximum
expected period difference scales quadratically with the sidereal period
for a given semi-major axis.
This behaviour follows the empirical relation
\begin{equation}
  \log_{10}\!\left|P_\text{synodic}-P_\text{sidereal}\right|
  = 2\,\log_{10} P_\text{sidereal} + \beta 
\end{equation}

\noindent where the slope is fixed and consistent across all simulated objects,
while the intercept $\beta$ depends solely on the semi-major axis of the target.

Both the maximum expected difference between the two periods and the number
of distinct intervals ($N$) that must be explored around the synodic solution
therefore depend on the semi major axis of the object ($a$). 
To capture this dependence, we fit these quantities as shown at the bottom panels of figure \ref{fig:window} and
derive the characteristic width of these intervals (in hours)
as 

\begin{subequations}\label{eq:interval_fits}
  \begin{align}
    N(a) &= 71.073\, e^{-1.21 a} + 2.528  \label{eq:interval_fits_a} \\
    \beta(a) &= 1.619\, e^{-0.338 a} - 5.069  \label{eq:interval_fits_b} \\
    W(a) &= 10^{\,2\log_{10} \psid+ \beta(a)}  \label{eq:Wa}
  \end{align}
\end{subequations}

\rev{The method still requires an accurate initial synodic period estimate, 
as an incorrect solution would propagate to the explored intervals
at the detriment of the ellipsoidal shape modeling.
An incorrect initial guess of the period would lead to a dephasing
of the lightcurve over the observations made around the reference epoch.
The final derived parameters, such as shape and 
spin axis coordinates, would be unreliable.
Therefore, the ability to retrieve an accurate synodic period 
estimate from the provided lightcurves can be seen as a minimum requirement for a \socca inversion.}

\subsection{Recovering true periods from aliases\label{ssec:pdm}}

We test the solutions flagged as aliases 
(\Cref{eq:trumpet}) to recover possible true periods. 
For cases with $k=1$, we repeat two \socca fits with trial
frequencies $f_\mathrm{obs} \pm \ffeat$. For cases with $k=2$, 
we repeat one \socca fit with a trial frequency $2f_\mathrm{obs}$.
These trial periods are computed as previously (\Cref{eq:aliasfunc}).

For each solution we compute the \shgg forward model using the \socca
fit parameters and determine the 
sub-Earth longitude (\Cref{eq:sep_long}).
We subtract the \shgg model from the data and phase fold the 
residuals over the sub-Earth longitude, producing a double-peaked lightcurve. 
We then use the 
the $\Theta$ statistic from \cite{1978ApJ...224..953S}
to select the true rotational frequency 
by phase dispersion minimization.

\begin{align}
\Theta &= \frac{S^2}{\sigma^2} \\
\sigma^2 &= \frac{\sum_{i=1}^{N} (x_i - \bar{x})^2}{N - 1} \\
S^2 &= \frac{\sum_{j=1}^{M} (n_j - 1) s_j^2}{\sum_{j=1}^{M} n_j - M}
\end{align}

\noindent where $\sigma^2$ is the total variance of the magnitudes $x_i$, and $S^2$ is the variance of the $M=100$ samples containing $n_j$ 
points with variance $s_j$. The trial period for which $\Theta$ is closest to unity is selected as the true rotational period.

\section{Validation on simulated data\label{sec:valid}}

\subsection{Simulating LSST-like data\label{ssec:sorcha}}

We use \sorcha \citep{2025JOSS...10.8145M}
to generate simulated lightcurves for 500 SSOs for each of the following
ten dynamical class: 
Aten, Amor and Apollo near-Earth asteroids, Mars-crossers
inner, middle and outer main belt,
Jupiter trojan
Centaurs,
and
KBOs. 
The orbital parameters correspond
to the actual parameters of the sampled objects, while the physical 
parameters are randomly drawn from physically motivated distributions accessed with 
the \rocks\footnote{\url{https://rocks.readthedocs.io}}
python client of 
\ssodnet
\citep{2023A&A...671A.151B}.
The absolute magnitude $H$ is sampled from a uniform distribution bounded by the 
minimum and maximum values of the observed $H$ distributions for each dynamical 
class. The spin period is drawn from the observed spin-period distribution of the 
entire SSO population, with the exception of near-Earth asteroids, for which a 
separate distribution is used, characterized by systematically shorter periods. 
The spin-axis coordinates \spinra\ and \spindec\ are sampled from the observed 
distributions of SSOs. The shape parameters \ab and \ac are taken from the 
\texttt{DAMIT} database of three-dimensional shape models, while 
enforcing the ordering $\ab < \ac$.

The only exception here are the \gone and \gtwo phase parameters and the \lsst colors required by \sorcha.
The phase parameters are selected according to the taxonomic class of the sampled objects,
using the scheme from \cite{2021Icar..35414094M}.
The \lsst colors ($u-r$, $g-r$, $i-r$, $z-r$, $y-r$) are
computed\footnote{\url{https://github.com/bcarry/ska}}
using the prototype spectra from 
\citet{2022A&A...665A..26M} 
using filters' transmission
curves\footnote{\url{https://svo2.cab.inta-csic.es/theory/fps/}} \citep{2012ivoa.rept.1015R,2020sea..confE.182R}.

Using these parameters, \sorcha
simulates\footnote{available to everyone using the
\texttt{ellipsoidalWithTerminator} lightcurve module we added
to \sorcha add-ons: \url{https://sorcha-addons.readthedocs.io}}
the photometric time series 
(Eq. \ref{eq:Hfgs}) and samples the resulting lightcurves as they would be 
observed by the \lsst\ over a ten-year baseline. From the simulated catalog, we 
retain only objects that have at least 50 observations in at least one filter, 
ensuring sufficient photometric coverage for a meaningful inversion. \rev{Out of the 5,000
total sampled objects, 2,816 were observed by the simulated \lsst\ survey and 2,207 were selected
 after having more than 50 observations.} The 
selected objects are then analyzed using the \socca\ model.

\subsection{Success rate\label{ssec:successrate}}
We apply \socca to 2,207 objects with a resulting success rate of 52.8\%
Of these 2,207 objects, 17 failed because \shgg did not converge, 
and four failed because \socca did not converge.
In addition 500 objects were flagged as having unreliable 
period estimates from the bootstrap criterion (\Cref{eq:bootstrap}).
A further 304 objects produced shapes for which \ab and \ac differ by less than $1\%$, 
indicating that the \ac axis of the ellipsoid is poorly
constrained. 

Additionally, 42 solutions returned non-physical values of \gone or \gtwo: 
For each model and each filter, we evaluate the number of successful solutions, 
defining a solution as valid when the \gone and \gtwo parameters are within
$> 5 \times 10^{-3}$ of the bounds
given in equations \ref{eq:cond_G1} to \ref{eq:cond_GG3}.
Solutions that do not satisfy 
these conditions are considered non-physical, as they approach the bounds of the parameter space,.
This behaviour in the optimization algorithm is a sign of poor constraints on the phase curve parameters 
from the data. \rev{Finally, 175 objects fail due to more than one of the failure modes listed above being true simultaneously,
as the conditions are not mutually exclusive.}
\rev{ The metrics above encompass the criteria for a successful \socca solution. They
quantify the requirements of sufficient phase and aspect angle coverage as well as adequate sampling of the lightcurve to retrieve
the rotational period.}
\rev{They can be implemented without the requirement of any previous
knowledge of the target SSO physical properties.}

Only a limited number of objects are recovered with the \hgg model, 
the dominant limitation being the unmodeled effect of shape. 
This limitation is mitigated with both
\shgg and \socca with the inclusion of the shape and shape-spin induced variability respectively.
As a consequence, we recover an additional $\sim$10--20\% of valid solutions per filter 
with \socca when compared to \hgg.

\begin{figure}[t]
  \input{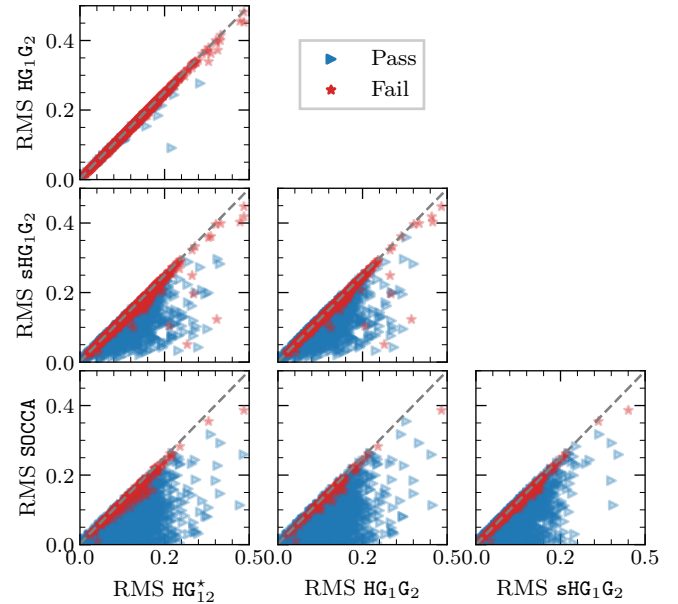}
  \caption{
    Comparison of the RMS between the different photometric models.
    The gray line represents the 1:1 line, where no improvement in the RMS is observed between models.
    The blue triangles depict the inversions
    for which the more complex model (with more free parameters)
    is justified by a statistically significant improvement
    in the residuals (\Cref{eq:f-test}).
    Conversely, the red stars represent those for which the 
    simplier model is more relevant.
  \label{fig:model_comparison}
  }
\end{figure}

\begin{figure*}[t]
  \input{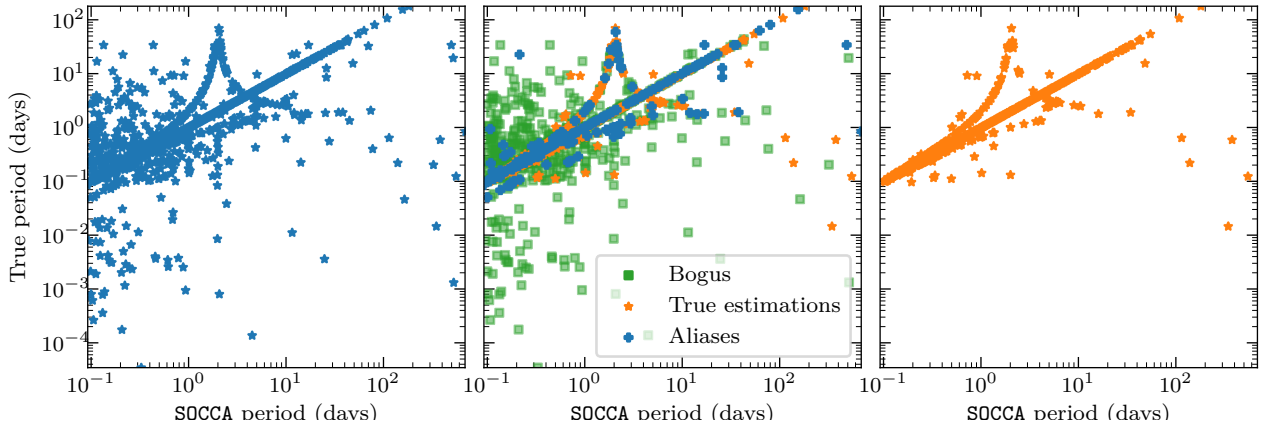}
\caption{Comparison between the simulated and retrieved periods for each simulated SSO. 
\textbf{Left:} Period estimates from the Lomb-Scargle periodogram. 
Both alias branches are visible above and below the 1:1 line, as well as the double-period 
harmonic below and parallel to the 1:1 line. 
\textbf{Center:} Application of our quality flags.
Green squares indicate low-scoring 
samples (\Cref{eq:bootstrap}), while blue crosses mark estimates flagged as aliases 
by the criterion of \Cref{eq:trumpet}. 
\textbf{Right:} Final estimates after removing bogus solutions and applying the phase 
dispersion minimization step. 
\label{fig:period_comparison}
}
\end{figure*}

\subsection{Computational cost and scalability}

\begin{figure}[H]
  \centering
  \input{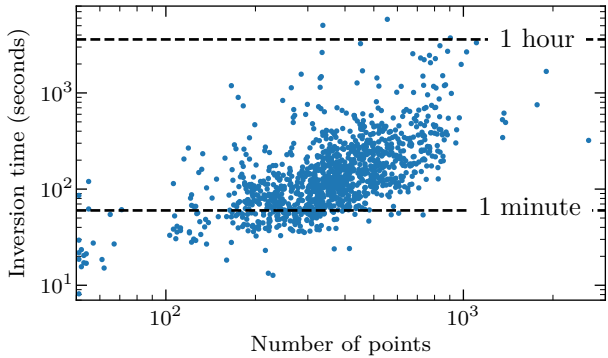}
  \caption{Inversion time as a function of the number of observations.
  The horizontal lines indicate 1 minute and 1 hour. Most solutions are
  obtained within a few minutes, with a clear dependence on the number
  of data points.}
  \label{fig:invtime}
\end{figure}

The application of \socca to the simulated sample of $2, 207$ objects allows
us to directly quantify the computational cost of the inversion. The
distribution of inversion times is shown in \Cref{fig:invtime}. The
median runtime is $\sim$157\,s per object, with 30\% of the solutions
converging in less than $\sim$87\,s and 90\% in less than $\sim$416\,s
on a single core.

The inversion time scales with the number of observations, as expected from
the least-squares minimization. This trend is visible in
\Cref{fig:invtime}, where the runtime increases approximately linearly
with the number of data points. Most solutions are obtained
within a few minutes, and the majority remain well below one hour.

Overall, the method scales well with the size of the dataset, and can be
applied to several thousand objects with modest computational resources.
This makes it directly applicable to current surveys such as \ztf, and
suitable for larger datasets expected from \lsst.

\subsection{Fit residuals}
As shown by \citet{2024A&A...687A..38C},
the \shgg model returns systematically lower RMS values 
than both \hg\ and \hgg, as it accounts for the variations in brightness induced 
by changes in the viewing geometry.
This behavior is confirmed here on \sorcha simulations,
with RMS values 
significantly reduced with respect to \hg\ and \hgg, although 
not fully reaching the expected observational uncertainty $\sigma$.
This 
indicates that while a substantial fraction of the photometric 
variability is captured by \shgg, an unmodelled source of scatter in the
lightcurve remains, due to the rotation of the object. 
This is captured by \socca, translating into a further
reduction of the RMS with respect to all previous models.
The \socca model provides much lower fit residuals with respect to previous models, with a mean
RMS$_{\socca}=0.07$, compared to RMS$_{\shgg}=0.11$ and RMS$_{\hgg} \approx \text{RMS}_{\hgs}=0.14$.
Similarly for the median reduced chi-squared, $\chi^2_{\nu, \socca} = 8.81$ while $\chi^2_{\nu, \shgg} = 61.9$
and $\chi^2_{\nu, \hgg} \approx \chi^2_{\nu, \hgs} = 108.9$

\socca introduces a total of $3\nfilter+6$ free parameters,
with $\nfilter$ the number of bands, 
significantly
more than the $2\nfilter$ parameters of
\hg or the $3\nfilter$ of \hgg.
We hence test whether the success \socca is due to an excessive 
number of parameters.
We adopt a hierarchical modeling strategy, in which the models are 
evaluated from the simplest to the most complex formulation.
The final choice 
is made using the F-test (\Cref{eq:f-test}), so that the most frugal 
model that provides a statistically significant improvement in the fit is 
retained.
Therefore, each increase in model complexity must
correspond to a measurable gain in descriptive power. Even under
this constraint, 88.5\% of the inversions tested here pass the 
F-test between all tested models and \socca
(\Cref{fig:model_comparison}).
The failure of \hgg and \hgs therefore reflects 
an incomplete physical description rather than over-parameterization. In this 
context, degrading the phase function formulation from \hgg to simpler 
variants such as \hgs does not address the underlying issue.
The appropriate response is to improve the physical
description of the model, as is done here with \socca.

\subsection{\Hmag, \gone, and \gtwo}\label{sec:G1G2}

\begin{figure}[b]
  \centering
  \input{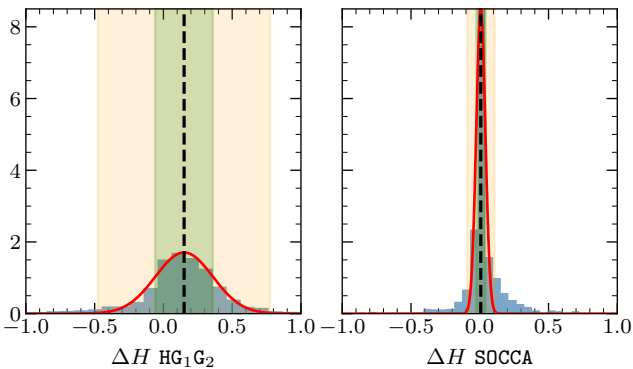}
  \caption{Difference between the simulated and computed \Hmag 
  for \hgg and \socca. Gaussian fits to the distributions are shown. 
  The \hgg residuals have a mean of 0.150 with $1\sigma = 0.209$ and 
  $3\sigma = 0.626$, while \socca\ shows a significantly tighter 
  distribution with a mean of 0.008, $1\sigma = 0.033$, and 
  $3\sigma = 0.099$.
\label{fig:H_comparison}
}
\end{figure}

We compare the absolute magnitude \Hmag and the phase function parameters 
\gone and \gtwo retrieved by \socca and the other photometric models with
the simulation inputs.
First, \socca retrieves precisely and accurately the absolute magnitude
(\Autoref{fig:H_comparison}). The average difference between the 
simulated and computed \Hmag is of 0.001\,mag only
and the scatter (i.e., the uncertainty)
is nearly three times smaller than with other models.
The offset seen in \hgg is due to the updated definition of \Hmag, see
\Cref{sec:model}.

Second, \socca also retrieves more precisely and accurately the phase function
parameters than other models (\Cref{fig:g1g2comp}).
The average difference between the simulated and estimated
\gone and \gtwo are about 0.03 on average, an
improvement of a factor three against previous models for \gone.
The dispersion is much reduced with \socca, about half that of \hgg, around
0.3 and 0.2 for \gone and \gtwo respectively.
This results in a much more informative \ggparams plane:
from \hgg to \shgg to \socca, the distribution becomes sharper,
allowing to clearly separate, e.g., the S- and C-type populations with \socca
(\Cref{fig:g1g2comp}).

\subsection{Spin axis orientation}

\begin{figure}[t]
  \centering
  \input{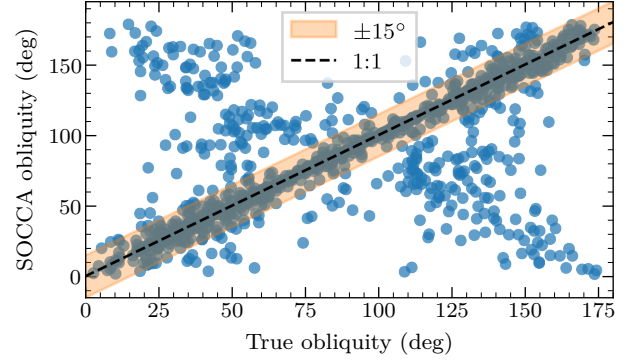}
  \caption{Input versus output spin axis obliquity of the sampled SSOs. The orange area
  represents the $\pm 15\degr$ we consider as a successful estimate}
  \label{fig:Psi_comparison}
\end{figure}

From the derived spin coordinates (\spinra, \spindec), 
we compute the obliquity of each target 
and compare it with the simulated obliquity
(\Cref{fig:Psi_comparison}).
68.5\% of the solutions lie within $15\degr$ of the true value.
Of the remaining 31.5\% of the failures, 25.5\%  
correspond to either incorrect period estimates 
or limited phase angle coverage ($<15\degr$). 
The erroneous obliquities are not randomly distributed 
and largely correspond to mirror spin solutions
(i.e., \spinra+180\degr, -\spindec).

\subsection{Rotational period}

To evaluate the performance of the period classification, we use the
$F_{\beta}$ score \citep{2005goutte}, which allows different weights to be
assigned to purity and completeness. A small value of $\beta$ places more
importance on purity, therefore we adopt $\beta = 0.25$.

\begin{figure}[t]
  \centering
  \input{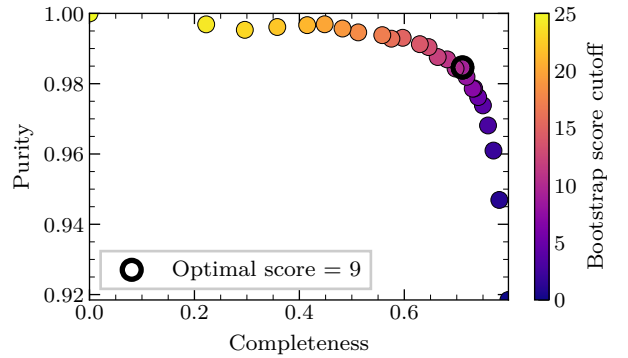}
  \caption{Purity and completeness as a function of the bootstrap score
  threshold used to identify bogus period estimates. The selected cutoff 
  of $9$ maximizes the $F_{\beta}$ score with $\beta=0.25$.}
  \label{fig:purity_v_completeness}
\end{figure}

To identify bogus period estimates we apply a cutoff on the bootstrap score
(\Cref{ssec:alias_period}).
The optimal threshold is found by maximizing $F_{\beta}$, resulting in a
selected cutoff of 9 (\Cref{fig:purity_v_completeness}).
We also identify the first-order aliases and the doubling of the true frequency, as illustrated
in \Cref{fig:period_comparison}, which are then re-evaluated with phase dispersion minimization
(\Cref{ssec:pdm}).
With these quality cuts and corrections, the purity of the true-period sample increases from
$60.4\%$ to $75.8\%$. 
The overall performance of the classification is summarized in
table \ref{tab:purity-completeness-periods}.

\begin{table}[H]
 \rev{\caption{Purity and completeness of each period class}
 \label{tab:purity-completeness-periods}}
\centering
\begin{tabular}{@{}cccccc@{}}
\toprule
      & Purity (\%) & Completeness (\%) & TP  & FP  & FN  \\ \midrule
True  & 75.8 & 83.7 & 1134 & 362 & 220 \\
Alias & 41.4 & 25.5 & 128  & 181 & 373 \\
Bogus & 41.6 & 84.1 & 296  & 415 & 56  \\ \bottomrule
\end{tabular}
\caption*{\del{Purity and completeness of each period class along with the number of
true positive (TP), false positive (FP) and false negative (FP) period estimates.
}}
\rev{\tablefoot{Purity and completeness of each period class along with the number Of
true positive (TP), false positive (FP) and false negative (FP) period estimates.}}
\end{table}

\subsection{Shape}

\begin{figure*}[t]
  \centering
  \input{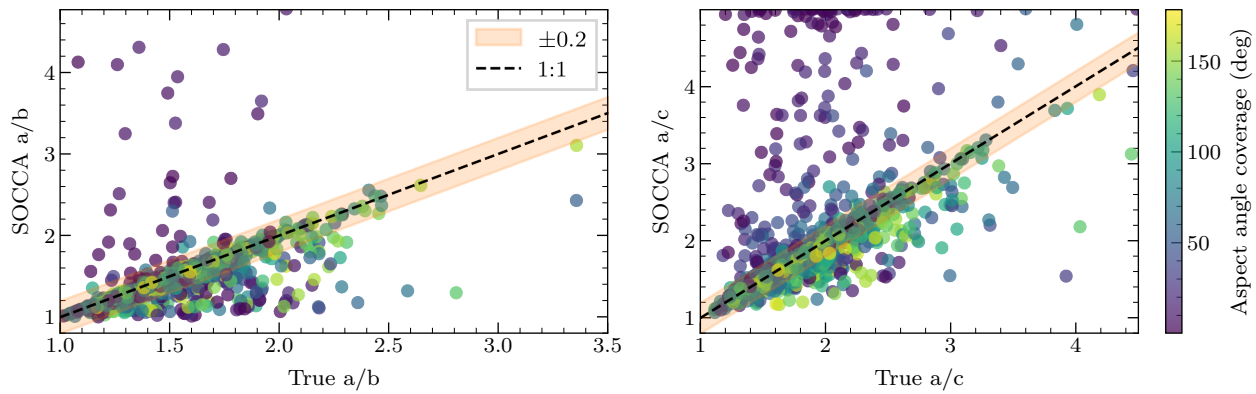}
  \caption{Comparison between the sampled and retrieved axis ratios
  $\ab$ and $\ac$. The orange area indicates the $\pm20\%$ region used to
  define successful solutions.}
  \label{fig:shape_copmarison}
\end{figure*}

We compare the axis ratios $\ab$ and $\ac$ obtained from the model with the
values used in the simulated sample. A solution is considered successful if
the retrieved value lies within $20\%$ of the sampled value.

Using this criterion, we obtain a success rate of $83.3\%$ for $\ab$ and
$67.4\%$ for $\ac$ (figure \ref{fig:shape_copmarison}). The higher success rate
for $\ab$ is expected, as this axis ratio is mainly constrained by the
spin-induced amplitude of the lightcurve. In contrast, $\ac$ is constrained
through changes in the aspect angle, which require observations spanning a
wide range of viewing geometries, in particular the aspect angle.
Such coverage is not always available in
the dataset, leading to a lower recovery rate.

\section{Validation on real data\label{sec:validate_eugenia}}

\begin{figure*}[t]
  \input{gfx/eugenia_model.pgf}
  \caption{Comparison between \socca, \hgg and \shgg model fitting for (45) Eugenia. The large panels
  show the model prediction versus the obervations obtained by \ztf in the g filter over the approximately 10 year baseline 
  and the smaller panels show the difference between the model prediction and the observations. On the top pair of panels the \socca - \hgg
  comparison is shown and on the bottom pair, we show the \socca - \shgg comparison. In the insets, the high 
  frequency signal due to the rotation of the asteroid is shown, which is modeled by \socca, but missed by both \hgg and \shgg, the later of which
  captures the shape variation.
    \label{fig:eugenia_fit}
  }
\end{figure*}

\begin{figure}[t]
  \input{gfx/eugenia_dense.pgf}
  \caption{Comparison between the densely sampled lightcurves of (45) Eugenia and the \socca model
  of the same SSO propagated to the dates of observation. \textbf{Top}: This lightcurve \citep{2016A&A...586A.108H} was captured near 
  the observational midpoint of the data used to build the \socca model and is therefore reproduced well.
  \textbf{Center}: This lightcurve \citep{2010Icar..210..635M}
  was captured under a certain object-observer orientation, leading to an offset in the magnitude axis due to the non-symmetry of the SSO.
  \textbf{Bottom}: This lightcurve was captured significantly earlier than the midpoint of observations (\textit{ibid}),
  leading to a visible dephasing between our model and the observations.
    \label{fig:eugenia_dense}
  }
\end{figure}

\noindent We evaluate the performance of the model on real observations.
We apply \socca to the asteroid (45) Eugenia, using photometric
measurements obtained by the Zwicky Transient Facility
\citep[\ztf,][]{2019PASP..131a8003M}.
The dataset consists of 265 photometric measurements acquired
between 2019-11-09 and 2026-02-23. The observations were retrieved from
\fink \citep{2021MNRAS.501.3272M} and are distributed in the $g$ and $r$ filters.

We compare the results with the \hgg and \shgg model. The \socca
solution produces significantly smaller residuals as seen in figure
\ref{fig:eugenia_fit}, indicating a better
description of the observed brightness variations.
More specifically, we retrieve a sidereal rotational period of
5.699,136 hours, compared to
5.699,152 hours obtained from 3D shape reconstruction using
light curves and adaptive-optics images \citep{2021A&A...654A..56V}.
Similarly, we derive spin axis coordinates of
$(\spinra, \spindec) = (119^{+2}_{-2}, -19^{+5}_{-5})\degr$
while the 3D shape reconstruction returns
$(\spinra, \spindec) = (121^{+2}_{-2}, -16^{+2}_{-2})\degr$,
only 4\degr away.
Finally, the reported shape axis ratios are $\ab = 1.319^{+0.08}_{-0.08}$ and $\ac = 1.826^{+0.11}_{-0.11}$, whereas \socca\ retrieves
$\ab = 1.366^{+0.03}_{-0.03}$ and $\ac = 2.165^{+0.16}_{-0.16}$.

We further validate the model using dense lightcurves of
(45) Eugenia available in the \texttt{DAMIT} database. The
\socca model is propagated to the epochs corresponding to the dense
lightcurve observations and directly compared with the measured
magnitudes.
In most cases, the model reproduces the dense lightcurves well, as seen 
at the top plot in figure \ref{fig:eugenia_dense}, where we compare with a dense lightcurve
taken in 2014 \citep{2016A&A...586A.108H}.
Deviations are expected because (45) Eugenia is not a perfectly
symmetric triaxial ellipsoid, while the model assumes such a geometry.
Certain viewing configurations therefore produce systematic offsets
between the predicted and measured brightness
(\Cref{fig:eugenia_dense}, middle).
Additional discrepancies arise from small differences in the
sidereal rotation period. The period derived by \socca differs from the
value reported in \texttt{DAMIT} by approximately $0.03$ seconds. Over long
time baselines this leads to a gradual phase drift.
As a result, lightcurves obtained many years before the ZTF
observations show a measurable phase shift. For example, the dense
lightcurve obtained in 1978 \citep{1980A&AS...40..257D} is offset by roughly $88\%$ of one rotation
cycle when compared with the propagated model.

\section{Discussion\label{sec:disc}}

The formulation presented here assumes a triaxial ellipsdoidal shape 
in principal-axis rotation. In this configuration, the model captures
the photometric variability induced by the shape and the rotation in tandem
with the phase angle variations.

This assumption can be modified without changing the overall structure of the
approach. For objects with irregular shapes,
or non principal-axis rotation, the description of the rotational
signal can be substituted with, e.g., 
cellinoid \citep{2014EM&P..112...73L},
tumbling rotation \citep{2001A&A...376..302K},
or 
full 3D model
\citep{2001Icar..153...24K}, still benefitting from the 
efficient search of initial conditions.

Another line of extension of \socca are binary systems,
which introduce additional variability through mutual eclipses and
occultations \citep{2009Icar..200..531S}.
These events alter the observed flux at specific
epochs and are not captured by a single-body description. Modeling such systems
requires accounting for the orbital configuration and the relative properties
of the components. Similarly, for the case of active asteroids and comets, 
the observed signal includes a contribution from the surrounding material,
which can vary independently of the rotational phase. 

Overall, the formulation presented here provides a baseline that is well suited
for the vast majority of SSOs. More complex situations can be addressed by extending
the model, at the cost of additional parameters.

Another interesting implementation of \socca is that of identifying candidates for 
follow-up observations. The number of candidates within SSOs is expected
to increase dramatically with the advent of \lsst, while the resources to conduct more detailed
dense lightcurve observing campaigns remains limited. The identification of interesting signatures
in the \socca model residuals (for example periodic deviations from the ellipsoid shape) can lead to
some kind of target prioritization.

\rev{We also note the fact that \lsst\ will observe a very big number of SSOs
(5.3 million discoveries) but with a relatively small median number of observations per object
\citep[between 23-234 detections, depending on dynamical class,][]{2025AJ....170...99K}. 
However, the model is not intended to operate exclusively on LSST photometry. An advantage of the method 
is its multi-filter design, which allows the combination of data from different surveys into unified lightcurves. \lsst\ is 
therefore used here primarily as a reference case and because it is expected to provide some of the highest quality photometric measurements 
for future combined datasets.}

\subsection{Model availability}

\socca is currently being integrated into the SSO scientific pipeline of the
\fink alert broker, and is expected to provide near real-time parameter
estimates for both \ztf and \lsst data streams, which will then be freely available
to the scientific community. 

The model is publicly available\footnote{\url{https://github.com/astrockers/socca-tune}}
and can be installed as a Python package. It is implemented as an extension
of the \texttt{phunk} library for phase-curve fitting\footnote{\url{https://github.com/astrockers/phunk}},
and follows the same interface. The user first defines a \texttt{PhaseCurve} object containing the photometric
data in array-like format. The arrays should include the reduced magnitude and associated uncertainties, the epoch
of each observation, the phase angle, and the corresponding filter.
From this, the sub-observer and sub-solar coordinates are computed. An initial
parameter vector is then constructed following the initialization steps layed out in 
\Cref{sec:init} and \Cref{sec:init:period}, after which the \socca model can be
fitted directly. The retrieved parameters are stored as attributes of the
phase-curve object. The fitting process is
illustrated here below:

\begin{lstlisting}[language=Python]
import phunk
from socca_tune.initialize import initialize

# Phase curve object from photometric data
pc = phunk.PhaseCurve(
    target="Eugenia",
    epoch=data["Date"],  # J2000
    phase=data["Phase"], # degrees
    mag=data["Reduced magnitude"],
    mag_err=data["Photometric errors"],
    band=data["Photometric filter ID"],
)

# retrieve coordinates
pc.get_ephems()

# initialize parameters
p0,_= initialize(pc, weights=pc.mag_err)

# run SOCCA fit
pc.fit(["SOCCA"], p0, weights=pc.mag_err)

# access fitted parameters
pc.SOCCA.alpha  # Spin axis right ascension
pc.SOCCA.delta  # Spin axis declination
pc.SOCCA.period # Sidereal rotational period
\end{lstlisting}

\section{Conclusions}\label{sec:conclusions}

We present \socca, a new approach to determine simultaneously the
spin, shape, and absolute magnitude (hence colors) of Solar system
objects (SSOs) from photometry such as produced nightly by large sky surveys.
\socca is a frugal approach, striving to minimize the number of 
free parameters. It can be seen as a compromise between 
simple phase curve fitting and derivation of full 3D shape models.
We propose a suite of preliminary steps to initialize the model, avoiding 
computationnally expensive approaches such as grid-search.

We validate \socca based on simulations of \lsst-like observations for
2,207 SSOs.
\socca finds a solution in $\approx$50\% of the cases. 
Among these, 
\socca systematically provides a better description 
of the observations, with significantly smaller residuals.
We also validate \socca on \ztf observations of (45) Eugenia. The parameters
determined with \socca matches those from the more complex 
3-D shape modeling approach. The magnitudes predicted by \socca
provide a good description of dense lightcurves, with the limitation of the 
symmetry imposed by the ellipsoidal shape.

The derived photometric parameters, 
the phase function coefficients \ggparams and 
absolute magnitude \Hmag in each filter, thus colors,
are more accurately determined
than with previous approaches (e.g., \hgg modeling).
This provides a more robust basis for taxonomic studies of large corpus of data.
The physical parameters introduced in \socca,
sidereal rotation period \psid,
spin-axis coordinates (\spinra, \spindec), and the
ratios of triaxial ellipsoid diameter (\ab, \ac),
are accurately retrieved by the algorithm.
This provides a new possibility to study the 
structure and diffusion of dynamical families, governed
by these properties through the Yarkovsky effect.

We designed \socca to be frugal and robust so it is practically
possible to apply it to very large samples, 
up to several $10^5$ SSOs, to shift from
individual studies to a population-level description of SSOs.
In the context of large surveys such as \lsst, such a model is
required to handle the volume of data.

\begin{acknowledgements}
  BC. and KOX. was supported by CNRS/INSU/PNP and CNES APR.
  We thank these programs for their support.
  This research used
  the \miriade \citep{2008LPICo1405.8374B},
  \ssodnet \citep{2023A&A...671A.151B}, and
  \topcat \citep{2005ASPC..347...29T} Virtual Observatory tools.
  It used the
  \astropy\footnote{\url{http://www.astropy.org}}
  \citep{astropy:2013,astropy:2018,astropy:2022},
  \sbpy\footnote{\url{https://sbpy.org/}}
  \citep{2019JOSS....4.1426M}, 
  \rocks\footnote{\url{https://github.com/maxmahlke/rocks}}
  \citep{2023A&A...671A.151B},
  and
  the Flat-Iron Institute \nifty
  \citep{Garrison_2024}
  python packages.
  This work was developed within the
  \fink community and made use of the \fink community broker resources. \fink
  is supported by LSST-France and CNRS/IN2P3. Thanks to all the developers and
  maintainers
\end{acknowledgements}

\bibliographystyle{aux/aa}
\bibliography{aux/ssodnet, aux/more}

\appendix

\section{Spheroid and ellipsoid models\label{app:socca}}

We detail here the computation of the shape-related \funcss fonction used
in \socca. First, let's recall the \funcs from \shgg \citep{2024A&A...687A..38C}:

\begin{eqnarray}
  \funcss &=& \funcs \nonumber\\
  &=& 2.5 \log_{10} \Big[ 1 - (1 - R) |\cos \Lambda| \Big] \label{eq:s}
\end{eqnarray}

\noindent where $0 < R \leq 1$ is the oblateness of the spheroid
\begin{equation}
	R = \frac{c(a+b)}{2ab} \label{eq:R}
\end{equation}

\noindent and the aspect angle $\Lambda$ is computed
from the
coordinates (\spinra,\spindec) of the spin axis
and the coordinates ($\alpha$,$\delta$)
of the asteroid as seen from the observer \citep{2021-riea}:
\begin{equation}
  \cos \Lambda = \sin{\delta} \sin{\spindec} +
  \cos{\delta}\cos{\spindec}\cos{(\alpha - \spinra)},
\end{equation}

The new \funcss of \socca is computed from the projection of
an ellipsoid on the plane of the sky:

\begin{eqnarray}
  \funcss &=& 2.5 \log_{10} \mathcal{A} \label{eq:s_socca}
\end{eqnarray}

To compute it, we define ($\varphi$, $90-\Lambda$)
and
($\varphi_s$, $90-\Lambda_s$) the planetocentric coordinates
of the subobserver and subsolar point on the target.
The unit vectors $\vec{e}$ and $\vec{s}$ to the observer and to the Sun
are thus:
\begin{eqnarray}
  \vec{e} &=&
  \begin{pmatrix}
    e_1 \\
    e_2 \\
    e_3 \\
  \end{pmatrix}
  =
  \begin{pmatrix}
    \sin\Lambda \cos\varphi \\
    \sin\Lambda \sin\varphi \\
    \cos\Lambda             \\
  \end{pmatrix}\\
  \vec{s} &=&
  \begin{pmatrix}
    s_1 \\
    s_2 \\
    s_3 \\
  \end{pmatrix}
  =
  \begin{pmatrix}
    \sin\Lambda_s \cos\varphi_s \\
    \sin\Lambda_s \sin\varphi_s \\
    \cos\Lambda_s               \\
  \end{pmatrix}
\end{eqnarray}

We then follow the formalism by 
\citet{1984Icar...57..443O}
to compute the projected surface area $\mathcal{A}$ of an ellipsoid
at opposition (phase angle $\gamma = 0$\degr) by 
introducing the matrix $\vec{Q}$ as
\begin{equation}
  \vec{Q} =
  \begin{pmatrix}
    1/a^2 & 0     & 0     \\
    0     & 1/b^2 & 0     \\
    0     & 0     & 1/c^2 \\
  \end{pmatrix}
\end{equation}

\begin{equation}
  \mathcal{A} = S_1 = \pi abc \left( \vec{e^t} \vec{Q} \vec{e} \right)^{1/2} \\
\end{equation}

where $^t$ indicates transposition, 
$a \ge b \ge c$ are the semi-major axes of the ellipsoid,
and with
\begin{equation}
  \vec{e^t} \vec{Q} \vec{e} =
  \frac{\sin^2 \Lambda\cos^2\varphi}{a^2}
  + \frac{\sin^2 \Lambda\sin^2\varphi}{b^2}
  + \frac{\cos^2\Lambda}{c^2} \label{eq:eQe}
\end{equation}

The generalisation to non-zero phase angle ($\gamma > 0$\degr) is based on
a partially illuminated ellipsoid
\citep{1984GeDed..17...87C}, which
apparent surface is delimited by the limb on one side and the
terminator on the other, hence:
\begin{eqnarray}
  \mathcal{A} &=& \frac{S_1 + S_2}{2} \\
  \textrm{where}\nonumber\\
  S_2 &=& \pi abc \left[
      \frac{ \vec{e^t} \vec{Q} \vec{s} }{ \left( \vec{s^t} \vec{Q} \vec{s}
    \right)^{1/2} }
    \right]
\end{eqnarray}

in which $\vec{s^t} \vec{Q} \vec{s}$ is computed with
equation \ref{eq:eQe} by replacing
($\varphi$,$\Lambda$) by
($\varphi_s$,$\Lambda_s$), and

\begin{eqnarray}
  \vec{e^t} \vec{Q} \vec{s} &=&
  \frac{\sin\Lambda \cos\varphi\sin\Lambda_s \cos\varphi_s}{a^2} \nonumber\\
  &+& \frac{\sin\Lambda \sin\varphi\sin\Lambda_s \sin\varphi_s}{b^2}\nonumber\\
  &+& \frac{\cos\Lambda \cos\Lambda_s}{c^2} \label{eq:eQs}
\end{eqnarray}

$S_2$ area is signed, to account for cases in which the phase angle
is above 90\degr,
i.e., the apparent surface is a lunula and not an ellipsoid.

The \funcss differs from the total flux from the projected
surface $-2.5 \log_{10} \mathcal{A}$ by a constant offset
$-2.5 \log_{10} \pi b c$, i.e., the magnitude
of the minimum projected surface.
This offset is incorporated in the absolute magnitude \Hmag, which then is
defined as the magnitude
of the object at 1\,au from both the Sun and the observer,
with a 0\degr phase angle,
seen from its equatorial plane
\textsl{and its prime meridian}.

Following the IAU recommendation \citep{2018CeMDA.130...22A} and
the recipes in \citet{2021-riea}, the subobserver longitude
(rotation phase $\varphi$) is computed as follow

\begin{equation}
  \tan(W - \varphi) = \frac{
    \cos \spindec \sin \delta - \sin \spindec \cos \delta \cos( \alpha-\spinra)
  }{
    \cos \delta \sin(\alpha - \spinra)
  }\label{eq:sep_long}
\end{equation}

with $W$ the longitude of the prime meridian at the time of observation,
computed as
\begin{eqnarray}
  W &=& W_0 + \dot{W} (t - t_0), \textrm{~with} \label{eq:W}\\
  \dot{W} &=& \frac{2\pi}{\psid}
\end{eqnarray}

\noindent from an initial angle $W_0$ at a reference epoch $t_0$.
The IAU recommendation is to take
$t_0$ as J2000 \citep{2018CeMDA.130...22A}.
It is, however, a transparent parameter and can be arbitrarily chosen at any epoch.
As the uncertainty on the prime meridian longitude $W$ grows linearly
with time (\Cref{eq:W}), leading to an uncertain orientation of the shape, 
it is more accurate to tie $t_0$ with the epochs of observations.
We typically taken the mid-epoch of observations to minimize the
dephasing issue linked with the reference epoch $t_0$.

\section{Model reparametrization\label{app:reparam}}

For practical numerical purposes such as
the stability of the inversion and 
the efficiency of the convergence,
most parameters are 
remapped into latent variables. 
This allows each physical parameter to remain within a valid range
while the optimized parameters are defined on $\R$.
We mainly use the following sigmoid function $\sig(x)$,
defined on \R and bounded to $]R,R+|C|[$,
and its inverse
function $\logit(x)$ to convert variables to/from latent variables.

\begin{eqnarray}
  \sig(x) &=& C + \frac{R}{1 + e^{-x}}\\
  \logit(p) &=& \log \frac{p}{1-p}, \quad p = \frac{x-C}{R}
\end{eqnarray}

The absolute magnitude \Hmag and the sidereal rotation period \psid\
being already defined on \R, we leave them untouched.
The axes ratios \ab and \ac are computed from the latent variables 
$u_{a/b}$ and $u_{a/c}$ using
\begin{eqnarray}\label{eq:shape_remap}
  \ab &=& 4 \, \sig(u_{a/b}) + 1 \\
  \ac &=& (5 - \ab) \, \sig(u_{a/c}) + \ab \\ 
\end{eqnarray}
\noindent with $R=1$ and $C=0$

\noindent This ensures that $1 < \ab < \ac < 5$.
The upper limit is somewhat arbitrary but chosen to
encompass extreme shapes. The condition between \ab and \ac is required
for principal axis rotator such as described by \socca.
Similarly to the axes ratio, we remap the \ggparams phase function parameters
into the latent variables $u_{\gone}$ and $u_{\gtwo}$ 
\begin{eqnarray}
   \label{eq:g1_remap}
  \gone &=& \sig_1(u_{\gone}) \\
   \label{eq:g2_remap}
  \gtwo &=& L + (U-L) \, \sig_2(u_{\gtwo})
\end{eqnarray}
\noindent with $\sig_1$ computed with $R = G_{\max} + |G_{\min}|$ and $C=G_{\min}$
and $L,U$ given by
\begin{eqnarray}
L &=& \max \left( G_{\min},\ a_1 \gone + b_1,\ a_3 \gone \right) \\ 
U &=& \min \left(G_{\max},\ a_2 \gone + b_2 \right)
\end{eqnarray}
\noindent where

\begin{eqnarray}
G_{\min} &=& -0.429 \qquad G_{\max} = 1.429 \\
a_1 &=& -3.9038 \qquad b_1 = -0.2445 \\
a_2 &=& -0.9635 \qquad b_2 = 1.0157 \\
a_3 &=& -0.4
\end{eqnarray}
\noindent from the constraints given by equations \ref{eq:cond_G1}-\ref{eq:cond_GG3}.
The initial rotation phase $W_0$, bounded to
$[-\pi/2, \pi/2]$,
is recovered from its latent representation $u_{W_0}$ as
\begin{equation}\label{eq:w0_remap}
    W_0 = \pi \, \sig(u_{W_0}) - \frac{\pi}{2}
\end{equation}
\noindent with $R=1$ and $C=0$

Finally, to account for the spherical geometry, 
the spin axis coordinates \spinra and \spindec are
expressed in latent Cartesian coordinates $(X, Y, Z)$, and 
expressed as
\begin{eqnarray}
  \label{eq:rho_remap}
  \rho &=& \sqrt{X^2 + Y^2 + Z^2} \\
  \label{eq:ra_remap}
  \spindec &=& \arcsin\frac{Z}{\rho} \\
  \label{eq:dec_remap}
  \spinra &=& \arctan \frac{Y}{X}
\end{eqnarray}

\section{Error calculation\label{app:reparam_err}}
\noindent The uncertainty on each fitted latent parameter is estimated from the
covariance matrix derived from the Jacobian of the optimal least squares solution.
Let $\mathbf{r}$ be the vector of residuals and $\mathbf{J}$ the Jacobian matrix
of partial derivatives evaluated at the solution $\mathbf{u}$ in latent space.
The covariance matrix of the parameters is approximated by

\begin{equation}
\mathbf{C} = \left(\mathbf{J}^{\mathsf T}\mathbf{J}\right)^{-1} \chi^2_\nu 
\end{equation}

\noindent where $\chi^2_\nu$ is the reduced chi-square of the fit,

\begin{equation}
\chi^2_\nu = 
\frac{\sum_i r_i^2}{N - M}
\end{equation}

\noindent with $N$ the number of observations and $M$ the number of fitted
parameters. The $1\sigma$ uncertainty on each latent parameter $u_i$ is then
obtained from the diagonal elements of the covariance matrix,

\begin{equation}
\sigma_{u_i} = \sqrt{C_{ii}} 
\end{equation}

\noindent Since the model parameters are optimized in latent space, the
uncertainties on the corresponding physical parameters are obtained through
standard error propagation. For a physical parameter $p = f(\mathbf{u_i})$
expressed as a function of $N$ latent variables $\mathbf{u_i}$, the variance is

\begin{equation}
\sigma_p^2 = \left(\sum_{k=1}^{N} \frac{\partial f}{\partial u_i} \sigma_{u_i} \right)^2
\end{equation}

\noindent Using the transformations defined in equation \ref{eq:shape_remap}, the propagated
$1\sigma$ uncertainties on the physical shape parameters $\ab$ and $\ac$ are

\begin{equation}
\sigma_{\ab} =
4\,\sig(u_{\ab})\left[1-\sig(u_{\ab})\right]\,
\sigma_{u_{\ab}}
\end{equation}

\begin{align}
\sigma_{\ac} &= \Big(
  \left[(1-\sig(u_{\ac}))\,\sigma_{\ab}\right]^2 \notag \\
  &\quad + \left[(5-\ab)\,\sig(u_{\ac})
  (1-\sig(u_{\ac}))\,\sigma_{u_{\ac}}\right]^2 \notag \\
  &\quad + 2(1-\sig(u_{\ac}))(5-\ab)\,\sig(u_{\ac}) \notag \\
  &\qquad \times (1-\sig(u_{\ac}))\,
  \sigma_{u_{\ac}}\sigma_{\ab}
\Big)^{1/2}
\end{align}\noindent with $R=1$ and $C=0$

\noindent Similarly, we propagate the uncertainties for \gone and \gtwo from equations \ref{eq:g1_remap}
and \ref{eq:g2_remap} as follows

\begin{eqnarray}
  \sigma_{\gone} &=& R \frac{e^{-u_{\gone}}}{(e^{-u_{\gone}} + 1)^2} \sigma_{u_\gone} \\
  \sigma_{\gtwo} &=& (U-L)(1-\sig(u_{\gtwo}))\sig(u_{\gtwo})\sigma_{u_{\gtwo}}
\end{eqnarray}
\noindent with $R$, $L$ and $U$ computed simililarly as in equations \ref{eq:g1_remap} and \ref{eq:g2_remap}

\noindent The uncertainty of the initial phase $W_0$ from equation \ref{eq:w0_remap} is
\begin{equation}
  \sigma_{W_0} = \pi \sig(u_{W_0})(1 - \sig(u_{W_0}))\sigma_{u_{W_0}}
\end{equation}
\noindent with $R=1$ and $C=0$

\noindent The uncertainty of the spin axis coordinates \spinra and \spindec is retrieved
form the the directional covariance of their latent counterparts given in equations \ref{eq:rho_remap}, \ref{eq:ra_remap} and \ref{eq:dec_remap}.
As \spinra and \spindec describe the direction of the spin vector, the magnitude $\rho$ has no physical meaning and is not
constrained by the photometric data. It is only used as it is necessary for the chosen reparametrization. Nonetheless it introduces 
some non-physical variance which should not be accounted for. Therefore, we compute the directional covariance of the $X$, $Y$ and $Z$ parameters as follows.

\begin{eqnarray}
\mathbf{v} &=& (X, Y, Z) \\
\mathbf{n} &=& \frac{\mathbf{v}}{\|\mathbf{v}\|} \\
\mathbf{P} &=& \mathbf{I} - \mathbf{n}\mathbf{n}^{T} \\
\mathbf{C}_{\mathrm{dir}} &=& \mathbf{P}\,\mathbf{C}_{XYZ}\,\mathbf{P}
\end{eqnarray}
\noindent where $\mathbf{I}$ is the idenity matrix.
The uncertainties of the latent parameters are given by the square of the diagonal element of 
the $3 \times 3$ matrix $\mathbf{C}_{\mathrm{dir}}$.

\noindent From these uncertainties we can retrieve the uncertainties of \spinra and \spindec as follows

\begin{eqnarray}
\frac{\partial \spindec}{\partial X} &=&
-\frac{XZ}{\sqrt{1-\frac{Z^2}{X^2+Y^2+Z^2}}\,(X^2+Y^2+Z^2)^{3/2}} \\
\nonumber \\
\frac{\partial \spindec}{\partial Y} &=&
-\frac{YZ}{\sqrt{1-\frac{Z^2}{X^2+Y^2+Z^2}}\,(X^2+Y^2+Z^2)^{3/2}} \\
\nonumber \\
\frac{\partial \spindec}{\partial Z} &=&
\left(\frac{1}{\sqrt{X^2+Y^2+Z^2}} -
\frac{Z^2}{(X^2+Y^2+Z^2)^{3/2}}\right) \nonumber \\ 
&\times& \frac{1}{\sqrt{1-\frac{Z^2}{X^2+Y^2+Z^2}}}
\end{eqnarray}

\begin{eqnarray}
\sigma_{\spindec} &=&
\sqrt{
\left(\frac{\partial \spindec}{\partial X}\sigma_X\right)^2 +
\left(\frac{\partial \spindec}{\partial Y}\sigma_Y\right)^2 +
\left(\frac{\partial \spindec}{\partial Z}\sigma_Z\right)^2
} \nonumber \\
& & {}+ 2\left(
\frac{\partial \spindec}{\partial X}
\frac{\partial \spindec}{\partial Y}
\sigma_X\sigma_Y
+
\frac{\partial \spindec}{\partial X}
\frac{\partial \spindec}{\partial Z}
\sigma_X\sigma_Z
\right. \nonumber \\
& & \left.
+
\frac{\partial \spindec}{\partial Y}
\frac{\partial \spindec}{\partial Z}
\sigma_Y\sigma_Z
\right)
\end{eqnarray}

\begin{eqnarray}
\sigma_{\spinra} &=&
\left(
\left(\frac{X}{X^2+Y^2}\sigma_Y\right)^2 +
\left(\frac{Y}{X^2+Y^2}\sigma_X\right)^2 \right) \nonumber\\
&-&\left.
\frac{XY}{(X^2+Y^2)^2}\sigma_X\sigma_Y
\right)^{1/2}
\end{eqnarray}

\section{\gone and \gtwo distributions per model}

-----------------------------------------------
\begin{figure*}[]
  \input{gfx/g1g2_per_model.pgf}
  \caption{%
    Distribution of \ggparams parameters for the different models.
  \textbf{Top:} Gaussian kernel density estimators (KDEs) of \ggparams for the 
  \hgg, \shgg, and \socca model. The contour encloses 68\% of the 
  integrated probability, corresponding to the $1\sigma$ level of a Gaussian 
  distribution. The white dash line represents the \texttt{G$_{12}^*$} phase parameter
  of the phase curves.
  \textbf{Bottom:}
  Distribution of the difference between the simulated and retrieved
  \gone and \gtwo for each model.
\label{fig:g1g2comp}
  }
\end{figure*}

\end{document}